\documentclass[authoryear,3p,review]{elsarticle}
\usepackage{lineno}
\usepackage{hyperref}   
\hypersetup{
  colorlinks   = true, 
  urlcolor     = cyan, 
  linkcolor    = cyan, 
  citecolor   = red 
}
\modulolinenumbers[5]

\journal{Advances in Space Research}

\usepackage[utf8]{inputenc}
\usepackage{graphicx}
\usepackage{amsmath}
\usepackage{bm}
\usepackage{siunitx}
\usepackage{longtable,tabularx}

\usepackage{booktabs,siunitx,multirow,enumitem}
\usepackage{subfig}
\usepackage[disable]{todonotes} 
\usepackage{xcolor,soul}
\colorlet{soulcyan}{cyan!30}
\colorlet{soulgreen}{green!70}
\colorlet{soulOrange}{orange!70}
\colorlet{soulOyellow}{yellow!60}
\colorlet{soulOblue}{blue!40}
\sethlcolor{soulOyellow}
\usepackage{arydshln}
\biboptions{sort&compress} 

\begin{document}

\begin{frontmatter}

\title{Dynamic-Programming-Based Failure-Tolerant Control for Satellite with Thrusters in 6-DOF Motion}
\author[]{Abdolreza Taheri%
\fnref{fn1}}
\ead{abdolreza.taheri@alum.sharif.edu}

\author[]{Nima Assadian\corref{cor1}%
\fnref{fn2}}
\ead{assadian@sharif.edu}
\address{Sharif University of Technology, Tehran, Iran}
\cortext[cor1]{Corresponding author, office phone: +98 (21) 6616 4607, fax: +98 (21) 6602 2731 }

\fntext[fn1]{Graduate Student, Department of Aerospace Engineering}
\fntext[fn2]{Associate Professor, Department of Aerospace Engineering}

\begin{abstract}
    In this paper, a dynamic-programming approach to the coupled translational and rotational control of thruster-driven spacecraft is studied. To reduce the complexity of the problem, dynamic-programming-based optimal policies are calculated using decoupled position and attitude dynamics with generalized forces and torques as controls. A quadratic-programming-based control allocation is then used to map the controls to actuator commands. To control the spacecraft in the event of thruster failure, both the dynamic programming policies and control allocation are reconfigured to cope with the losses in controls. The control allocation parameters are adjusted dynamically to ensure the satellite always approaches the target from the side with two operative thrusters to achieve a stable control. The effectiveness of the proposed dynamic programming control is compared with a Lyapunov-stable control method, which shows that the proposed method is more fuel-efficient in tracking the same path.
\end{abstract}


\end{frontmatter}


\pagebreak 

\section*{Nomenclature}
{\renewcommand\arraystretch{1.0}
\noindent\begin{longtable*}{@{}l @{\quad=\quad} l@{}}
$\bm{C_X^Y}$ & rotation matrix from coordinate frame X to Y \\ 
$\bm{\hat{e}}$ & Euler axis unit vector\\
$\bm{F}^{B},\bm{F}^{R}$ & total external force vector in body and RSW frames\\ 
$\bm{F}^\ast$ & optimal force vector\\ 
$\bm{H}$ & thrust distribution matrix \\ 
$\bm{H}_c$ & thrust distribution matrix for 2-DOF channel \\ 
$\tilde{\bm{H}},\bm{Q},\bm{R}$ & cost weighting matrices\\ 
$\bm{J}$ &  inertia matrix \\ 
$J^\ast$ & optimal cost \\ 
$\bm{M}^B$ & total external torque vector in body frame \\ 
$\bm{M}^\ast$ & optimal moment vector\\ 
$\bm{q}$ & unit quaternion vector\\
$\bm{R}$ & position vector of the target satellite \\ 
$\bm{t_{\textrm{on}}}$ & thruster-on pulse duration \\
$\bm{u}$ & vector of the thrusters' forces \\ 
$\bm{u}_c$ & vector of the thrusters' forces for 2-DOF channel \\
$\tilde{\bm{u}}$ & dynamic programming control vector \\ 
$\bm{u}_\textrm{control}$ & solution to the quadratic-programming problem \\ 
$\bm{u}_\textrm{actual}$ & pulse-modulated thruster commands \\ 
$u_\textrm{max}$ & maximum effective thrust force during the control period\\
${u}_\textrm{on}$ & thrusters' fixed thrust force\\
$\bm{\hat{u}}$ & Lyapunov-based thrusting strategy \\
$\bm{V}$ & velocity vector of the target satellite \\ 
$\bm{\bar{w}}_F,\bm{\bar{w}}_M$ & ideal controls of the Lyapunov-based method \\
$\bm{x}$ & dynamic programming state vector \\ 
$\theta$ &  rotation angle about the Euler axis\\ 
$\bm{\rho}$ & relative position vector \\ 
$\bm{\sigma}$ & modified Rodriguez parameters vector \\ 
$\bm{\omega}$ & angular velocity vector in body frame \\

\end{longtable*}}

\section{Introduction}
The development of novel and efficient methods for controlling satellites in coupled six-degree of freedom (6-DOF) relative motion enhances the effectiveness of many orbiting satellite missions. Close proximity operations, rendezvous and docking, formation flying, and on-orbit servicing are instances of such technologies that offer cost-effective and flexible missions to be conducted by flying multiple satellites in relative motion, delivering capabilities that would not be achievable with a single large spacecraft \citep{Leitner2004,brown2006fractionated}. There are many advantages to employing small satellites in formation instead of a large spacecraft, namely overcoming the size and mass limitations of current launch vehicles, speeding up mission development time, and reducing the costs. Moreover, failure in a monolithic system may compromise the entire mission, whereas the distributed architecture of small, less-complex satellites in formation may continue to function with degraded level in performance in case one of the satellites encounters failure \citep{Sabol2001, mazal2014optimal}.

Applications of autonomous formation flying and close proximity operations of satellites have been reviewed in some previous studies \citep{Rupp2007,Leitner2004,Sandau2010,scharf2004survey}. In some applications, the satellites are required to fly in millimeter precision \citep{Rupp2007,inamori2011compensation}. In addition to reaching high control accuracy, the fuel consumption of the maneuvers must be kept to a minimum in order to increase the operational life of the satellites \citep{yoo2013spacecraft,mazal2014optimal}. This is especially important in the case of satellites having on-off thrusters where position and attitude dynamics are coupled \citep{Yang2003,Heydari2013, li2016state, Singhose2006,VanderVelde1983,yoshimura2018optimization}. Previous works by \citet{Yang2003} and \citet{Curti2010} deal with the problem of controlling thruster-driven satellites in 6-DOF. More specifically, these studies address selecting the best set of actuator on-off commands in order to achieve stability in the controlled motion. 

In a study of 156 on-orbit spacecraft failures from 1980 to 2005, thruster failures (including electric propulsion systems) collectively accounted for 24\% of all attitude and orbit control subsystem failures \citep{Tafazoli2009}. These failures are likely to occur at any time during the mission lifetime of the spacecraft, and the spacecraft may become unrecoverable if not handled by a failure recovery system or through redundancy. Additional thrusters can always be incorporated in a spacecraft to replace the failed thrusters when failures have occurred, as is the case with large spacecraft that usually have cost budgets in order of billions of dollars. However, for the new generation of small satellites, the cost of utilizing redundant actuators outweighs the total cost of failure \citep{Sarsfield1998}, and significantly increases the design mass and volume of the system as well. In this regard, a failure-tolerant system would be the ideal solution for handling under-actuation in small satellites. Controlling thruster-driven spacecraft in 6-DOF under thruster-failures has been studied by \citet{Pong2010,Tavakoli2018,tavakoli2018actuator}, and there is extensive research on fault-tolerant position-only \citep{huang2015analytical} or attitude-only control \citep{cao2013fault,cao2014minimum,fazlyab2016adaptive,sun2019constrained}, mostly considering spacecraft actuated with reaction wheels.

As a representative model of fully-actuated all-thruster spacecraft we consider the Synchronized Position Hold, Engage, Reorient, Experimental Satellites (\textit{SPHERES}) \citep{Miller2000,Saenz-Otero2002} in our analysis for several reasons. First, the \textit{SPHERES} is controlled with twelve thrusters that is the minimum number of actuators necessary for full-actuation. When any of the thrusters become inoperative, the satellite becomes under-actuated, making it a suitable candidate for implementing failure-tolerant control. In addition, its fast rotational maneuvering capability and the jittering induced by the thrusters must be handled by delicate control methods to reach the objectives of precision and optimal fuel consumption. Previous works have also been conducted on fault-tolerant control of \textit{SPHERES}. \citet{Pong2010} studied a Model Predictive Control (MPC) approach to controlling \textit{SPHERES}. While the MPC method proposed by \citet{Pong2010} assumes that the faults have already been identified through an autonomous fault detection and isolation process, \citet{Tavakoli2018,tavakoli2018actuator} applied neural networks to identify thruster faults based on the predicted behavior of the system and to continuously update the spacecraft model in MPC. This way, the need for a separate fault detection and isolation (FDI) system is eliminated.  Furthermore, in the geometry of \textit{SPHERES}, each thruster generates forces and torques only in one direction. This allows the state equations to be decomposed into independent 2-DOF systems, which makes the model suitable for dynamic programming analysis.

Dynamic programming is a powerful tool yet practically difficult to be utilized for problems in high dimensional state and action spaces such as satellite 6-DOF motion. The computational difficulties that result from this so-called "\textit{curse of dimensionality}" are so challenging that the literature on applying dynamic programming to spacecraft control is currently limited to one study on electromagnetic satellite actuation \citep{Eslinger2013}. 

In this paper, a dynamic-programming-based approach to coupled position and attitude control of all-thruster spacecraft is developed. Dynamic programming is used for computing the optimal generalized forces and torques as control policies. An efficient dynamic programming algorithm is also proposed for the time-invariant dynamical system, which speeds up the computations using vectorized operations supported by many modern processors. Then, the optimal policies from the dynamic programming approach are mapped to thruster on-off commands using a control allocation system. The performance of the proposed dynamic-programming-based control is compared with the Lyapunov-based thrusters' selection approach previously developed by \citet{Curti2010} for fully-actuated satellites having a generic thrusters' geometry. The Lyapunov-based approach is essentially a reference trajectory following method with mathematical proof for asymptotic stability, provided that certain conditions are met. The problem of controlling the under-actuated satellite is then investigated by adjusting the dynamic programming parameters and the weighting variable of the control allocation system. Effective switching between position control and attitude control causes the satellite to approach the target from the side with fully operative thrusters to achieve a stable control. 

The structure of this paper is as follows. The governing equations of motion are introduced in Section~\ref{sec:dynamics}. Section~\ref{sec:dynamic_prog} presents the dynamic programming method and the proposed algorithm. Section~\ref{sec:control_aloc} gives details on the quadratic-programming-based thruster mapping and the pulse-modulation technique. Section~\ref{sec:failure} presents the proposed methodology for control in failure conditions, and Section~\ref{sec:lyap} is devoted to a brief overview of the Lyapunov-based control. Section~\ref{sec:simul} presents the simulation results, demonstrates the advantages of the proposed control system, and offers a discussion of the analysis. Section~\ref{sec:conclusion} concludes this paper.

\section{Satellite Dynamics Modelling} \label{sec:dynamics}

In this section, the equations of motion for satellite relative motion are presented. The simulations for both the dynamic-programming-based and Lyapunov-based control in Section~\ref{sec:simul} are conducted using the coupled translational and attitude dynamics described in this section.  

\subsection{Rotational Dynamics} \label{sec:att_dynamics}

\begin{figure}[!htp]
    \centering
    {\includegraphics[width=0.65\columnwidth]{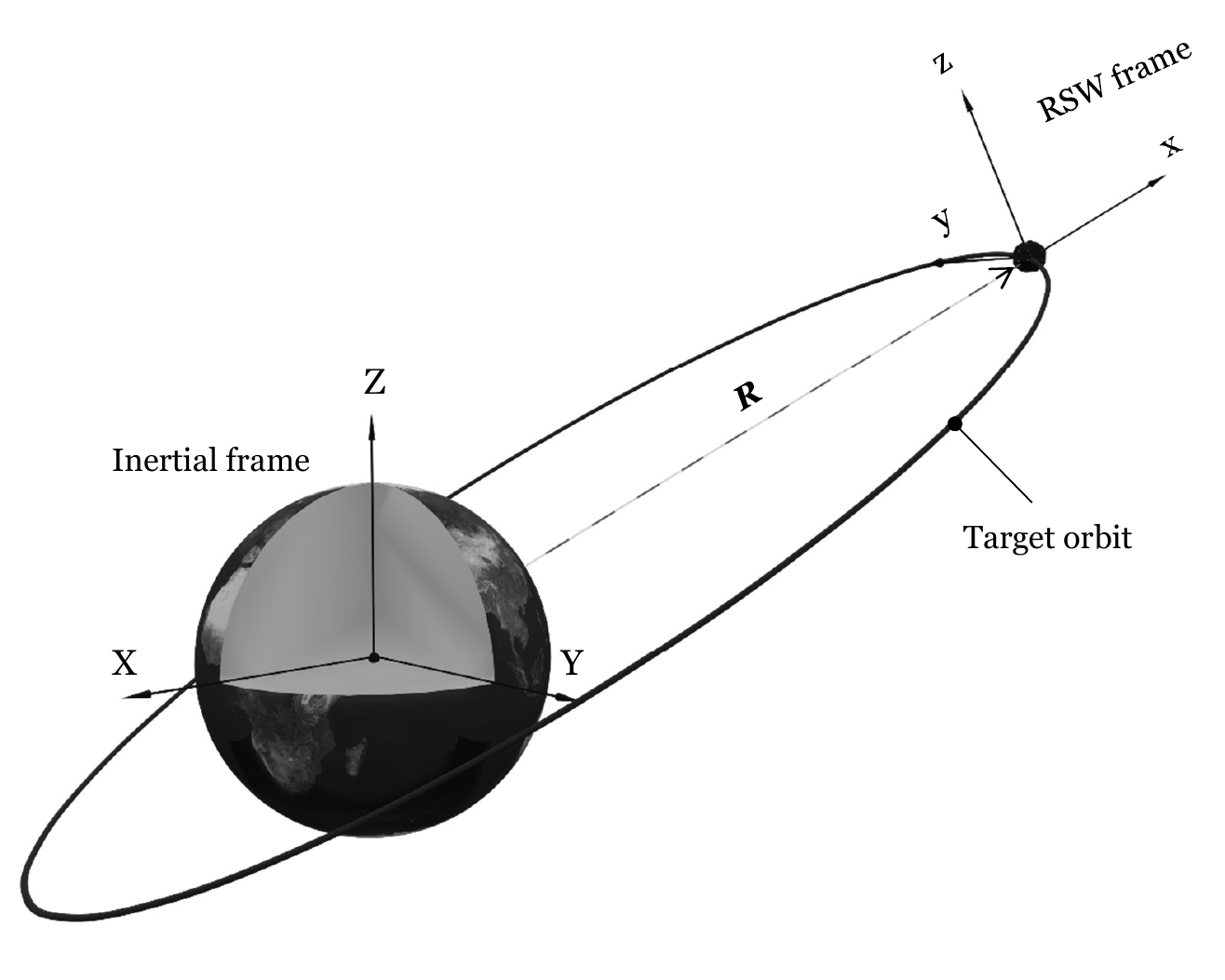}}
    \caption{The Earth-Centered Inertial (ECI) and RSW reference frames}
    \label{fig:CoordinateSystems}
  \end{figure}

The satellite attitude is described by a quaternion vector $\bm{q}$ defined as \citep{Wie2008}
\begin{equation}
    \label{eq:quat}
    {\bm{q}} = {\begin{bmatrix}
        {{{\bm{q}}_{13}}}&{{q_4}}
        \end{bmatrix} ^T} = {{\begin{bmatrix}
        {{\bm{\hat e}}\sin \frac{\theta }{2}} & {\cos \frac{\theta }{2}}
        \end{bmatrix}} ^T}
\end{equation}
where $\bm{\hat{e}}$ represents the Euler axis unit vector and $\theta$ is the rotation angle about the Euler axis. This quaternion vector represents the rotation from the Earth-Centered Inertial coordinate system (ECI) to the body frame. In the ECI coordinate frame shown in \figurename{~\ref{fig:CoordinateSystems}}, the origin is on Earth's center, the X-axis points toward the vernal equinox while the Z-axis is in the direction of Earth's rotation axis, and Y-axis completes the right-handed coordinate system. Subsequently, the kinematic equations for the satellite are written as Eq.~\eqref{eq:kinematic}

\begin{align}
    \label{eq:kinematic}
    &{\bm{\dot q}} = \frac{1}{2}{\bm{\Omega (\omega )q}} \\
    &{\bm{\Omega (\omega )}} = \left[
         {\begin{array}{ccc;{2pt/2pt}c}
        0 & {{\omega _3}} & { - {\omega _2}} & {{\omega _1}} \\
        { - {\omega _3}} & 0 & {{\omega _1}} & {{\omega _2}} \\
        {{\omega _2}} & { - {\omega _1}}& 0 &{{\omega _3}}\\ \hdashline[2pt/2pt]
        { - {\omega _1}}&{ - {\omega _2}}&{ - {\omega _3}}& 0
        \end{array}} \right] \nonumber
\end{align}
where $\bm{\omega}$ is the angular velocity vector of the body frame with respect to inertial frame. Moreover, the rotational dynamics of the satellite as a rigid body are described by the general form of Euler's rotation equation \citep{Wie2008}
\begin{equation}
    \label{eq:rotdyn}
    \dot{\bm{\omega}}=\bm{J}^{-1}\left(-\bm{\omega}\times\bm{J\omega}+\bm{M}^B\right)
\end{equation}
where $\bm{J}_{3\times3}$ denotes the inertia matrix and $\bm{M}^B$ is the total external torque vector generated by the thrusters. It should be noted that the other perturbations effects are negligible in comparison to the thrusters' forces and moments.

\subsection{Relative Translational Dynamics} \label{sec:pos_dynamics}

The relative position between the chaser and its target $\bm{\rho} = \left[{\delta x} \ \ \delta y \ \ \delta z\right]^T$ is expressed in the RSW frame shown in \figurename{~\ref{fig:CoordinateSystems}}. In this co-moving frame, the x-axis is aligned with the position vector $\bm{R}$ from the Earth center to the target satellite, z-axis is normal to the orbital plane in the direction of the orbital angular momentum, and y-axis completes the right-handed coordinate system. The translational dynamics are given by the linearized equations of relative motion in orbit \citep{Curtis2013}
    \begin{equation}
        \label{eq:CWlin}
    \ddot{\bm{\rho}}=f(\bm{\rho},\dot{\bm{\rho}},t)=\begin{bmatrix}\left(\frac{2\mu}{R^3}+\frac{h}{R^4}\right)\delta x-\frac{2(\bm{V}.\bm{R})h}{R^4}\delta y+\frac{2h}{R^2}\delta\dot{y}\\\left(\frac{h}{R^4}-\frac{\mu}{R^3}\right)\delta y-\frac{2(\bm{V}.\bm{R})h}{R^4}\delta x+\frac{2h}{R^2}\delta\dot{x}\\-\frac{\mu}{R^3}\delta z\\\end{bmatrix}+\frac{1}{m}\begin{bmatrix}F_x^R\\F_y^R\\F_z^R\\\end{bmatrix}
    \end{equation}
    where $\bm{R}$ and $\bm{V}$ represent the position and velocity vectors of the target orbit. These vectors vary with time and are updated continuously using the orbital elements \citep{Curtis2013}. $\bm{F}^R$ is the external force vector resulting from the thrusters' reactions in the RSW frame of reference and is expressed in terms of body forces $\bm{F}^B$ as
    \begin{equation}
        \label{eq_6}
        \bm{F}^R=\bm{C}_B^RF^B
    \end{equation}
    where $\bm{C}_B^R$ denotes the rotation matrix from the body frame of reference to RSW frame of reference, which is obtained from Eq.~\eqref{eq:rotmat}, the terms $\bm{C}_I^R$ and $\bm{C}_B^I$ represent the rotation matrices from the inertial frame to RSW, and from body to inertial frame, respectively.
    
    \begin{align}
        \label{eq:rotmat}
            &\bm{C}_B^R=\bm{C}_{I}^R\bm{C}_B^{I}\\
            &\bm{C}_B^{I}=q_4-(\bm{q}^T\bm{q})\bm{I}_{3\times3}+2\bm{q}\bm{q}^T-2q_4(\bm{q}^\times)\cdot\bm{q} \nonumber\\
        &\bm{C}_{I}^R=\left[\begin{matrix}\hat{\bm{R}}&\hat{\bm{S}}&\hat{\bm{W}}\\\end{matrix}\right]^T \nonumber\\
            &\hat{\bm{R}}=\frac{\bm{R}}{\left|\bm{R}\right|}\ , \ \ \
            \hat{\bm{W}}=\frac{\bm{R}\times\bm{V}}{\left|\bm{R}\times\bm{V}\right|}\ , \ \ \
            \hat{\bm{S}}=\hat{\bm{W}}\times\hat{\bm{R}} \nonumber
    \end{align}

\subsection{\textit{SPHERES} Control Forces and Moments} \label{sec:SPH_dynamics}

\begin{figure}[!htp]
    \centering
    {\includegraphics[width=0.75\columnwidth]{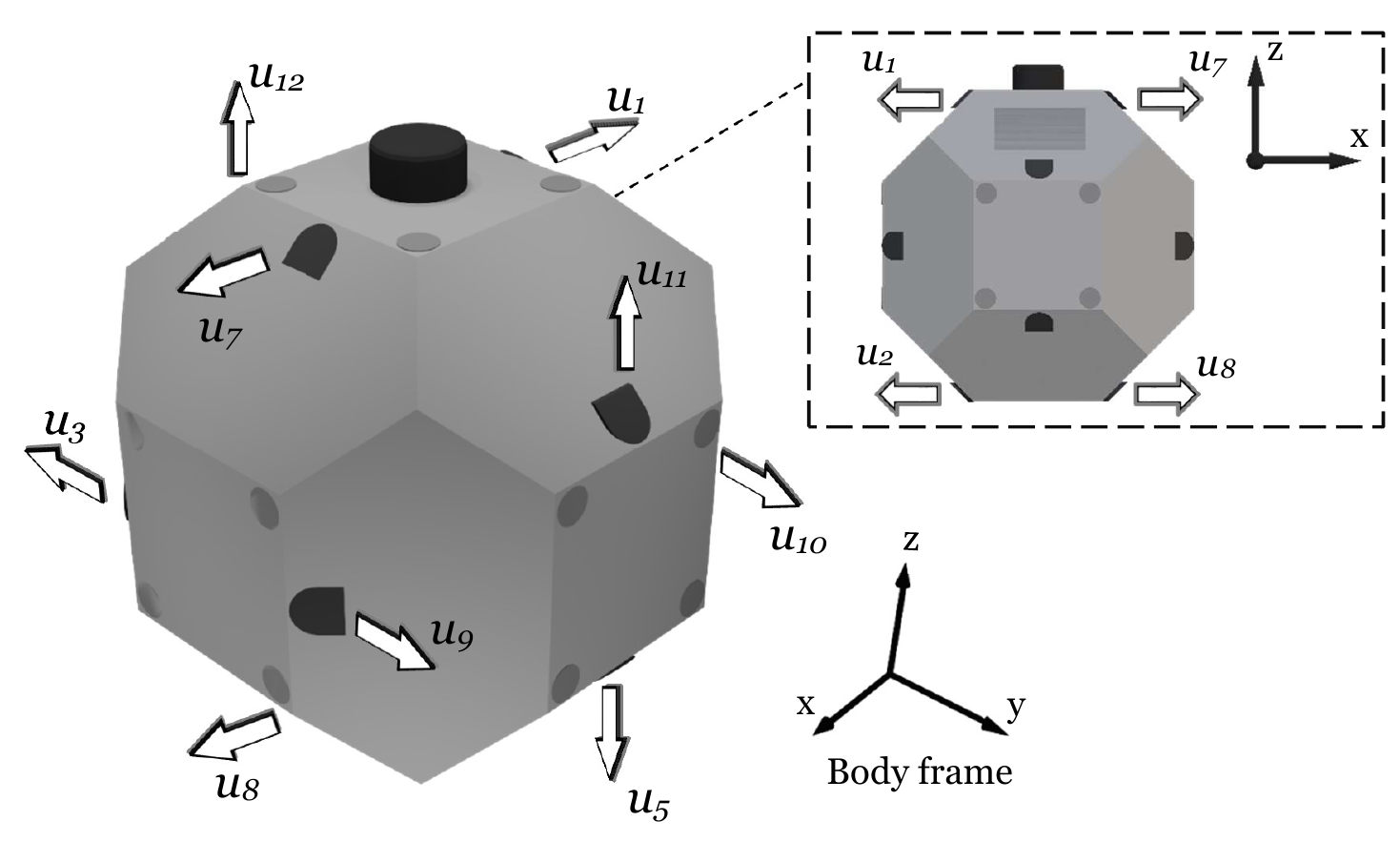}}
    \caption{Thrusters' geometry in \textit{SPHERES} and view of the 2-DOF channel \textit{x}}
    \label{fig:SPHERESthrusters}
  \end{figure}

Twelve cold-gas thrusters provide full 6-DOF actuation for the \textit{SPHERES} satellite, as shown in \figurename{~\ref{fig:SPHERESthrusters}}. The thrusters are arranged as six back-to-back pairs that produce forces in opposite directions. Given a vector of thrusters' exerted forces $\bm{u}=\left[u_1,u_2,\ldots,u_{12}\right]^T$, Eq.~\eqref{eq:bodyfm} can be used to determine the generated body forces and torques from thruster on-off commands.
\begin{equation}
    \label{eq:bodyfm}
\left[\begin{matrix}\bm{F}^B\\\bm{M}^B\\\end{matrix}\right]=\bm{Hu}=\left[\begin{matrix}\bm{H}_F\\\bm{H}_M\\\end{matrix}\right]\bm{u}
\end{equation}
Furthermore, the values of the thrust distribution matrices $\bm{H}_F$ and $\bm{H}_M$ are indicated in Eq.~\eqref{eq:Hmat}, which are derived from the geometry of the satellite's thrusters in \figurename{~\ref{fig:SPHERESthrusters}}. Table~\ref{tab:props} lists the thruster lever arm $d$ along with mass and inertia properties of the satellite \citep{Hilstad2010}.

\begin{table}[htbp]
    \centering
    \caption{\textit{SPHERES} specifications \citep{Hilstad2010}}
      \begin{tabular}{lc}
        \toprule
        Quantity & Value \\
        \midrule
        Total mass ($m$) & 4.16 \si{\kilo\gram} \\
        Thrust force ($u_\textrm{on}$) & 0.12 \si{\newton} \\
        Thruster lever arm ($d$) & 9.65 \si{\centi\meter} \\
        Inertia matrix ($J$) & $10^{-2}\left[\begin{matrix}2.3&0.01&-0.03\\0.01&2.42&-0.003\\-0.03&-0.003&2.14\end{matrix}\right]$ \si{\kilo\gram}.\si{\meter\squared} \\
      \bottomrule
    \end{tabular}
    \label{tab:props}
  \end{table}

  {
  \begin{equation}
    \label{eq:Hmat}
    \begin{array}{rl}
\bm{H}_F= & \begin{bmatrix}1&1&0&0&0&0&-1&-1&0&0&0&0\\0&0&1&1&0&0&0&0&-1&-1&0&0\\0&0&0&0&1&1&0&0&0&0&-1&-1\\\end{bmatrix} \\
    \\
\bm{H}_M= & \begin{bmatrix}0&0&0&0&d&-d&0&0&0&0&-d&d\\d&-d&0&0&0&0&-d&d&0&0&0&0\\0&0&d&-d&0&0&0&0&-d&d&0&0\\\end{bmatrix} \\
    \\
\end{array}
\end{equation}
}
As can be inferred from the thrust distribution matrices, the thrusters are mounted in a way that every four thrusters exert forces solely along one body-fixed axis and produce torques about another. In this line of thought, the system described by Eqs.~\eqref{eq:bodyfm} and \eqref{eq:Hmat} can be decomposed into three separate 2-DOF systems, one along each of the body axes that we refer to as \textit{channels}. A similar 2-DOF system decomposition analysis has been used by \citet{Jewison2014}. Figure \ref{fig:SPHERESthrusters} and Eq.~\eqref{eq:channel} illustrate one of such 2-DOF systems that is associated with the force acting along the x-axis and torque about y-axis.

\begin{equation}
    \label{eq:channel}
{\underbrace{\begin{bmatrix}
1&1&-1&-1\\
d&-d&-d&d
\end{bmatrix}}_{\bm{H}_c}}
\underbrace{\begin{bmatrix}\,u_1\,\\\,u_2\,\\\,u_7\,\\\,u_8\,
\end{bmatrix}}_{\bm{u}_c}
= \begin{bmatrix}F_x^B\\M_y^B\end{bmatrix}
\end{equation}

\section{Satellite 6-DOF Control} \label{sec:Control}
\subsection{Dynamic Programming Control Law Design} \label{sec:dynamic_prog}
In this section, we aim to find an optimal policy ${\tilde{\bm{u}}}^\ast=f(\bm{x}_{\left(k\right)})$ that minimizes the following quadratic cost function over $N$ equally spaced time increments from $t_0 = 0$ to $T_f$
\begin{equation}
    \label{eq:costfun}
    J=\frac{1}{2}\bm{x}_{\left(N\right)}^{T}\tilde{\bm{H}}\bm{x}_{\left(N\right)}+
    \sum_{k=0}^{N-1}{g_D(\bm{x}_{\left(k\right)},{\tilde{\bm{u}}}_{\left(k\right)})}
\end{equation}
\begin{equation}
    \label{eq:gD}
    g_D(\bm{x},\tilde{\bm{u}})=\frac{1}{2}\left(\bm{x}^{T}\bm{Q}\bm{x}+{\tilde{\bm{u}}}^{T}\bm{R}{\tilde{\bm{u}}}\right)
\end{equation}
where $\bm{x}$ and $\tilde{\bm{u}}$ are the state and control vectors, respectively, $\tilde{\bm{H}}$ and $\bm{Q}$ are real symmetric positive semi-definite $n\times n$ matrices and $\bm{R}$ is a real symmetric positive definite $r\times r$ matrix. The optimal value function is then computed iteratively by stepping backwards from the final stage using the functional equation \citep{Kirk2012}
\begin{equation}
    \label{eq:func}
    \begin{aligned}
    & J_{N-K,N}^*(x_{N-K}) = & & \underset{\tilde{u}_{N-K}}{\text{ min}} \left\{g_D(x_{N-K},{\tilde{u}}_{N-K})+J_{N-(K-1),N}^*(f_D(x_{N-K},{\tilde{u}}_{N-K}))\right\}\\
    & & & \text{subject to} \quad x_{k+1}=f_D(x_k,{\tilde{u}}_k)
    \end{aligned}
 \end{equation}
    where $J_{N-K,N}^\ast$ denotes the optimal cost in the last $K$ stages and the discrete solution  $x_{k+1}$  to the time-invariant system described by the state equation $\dot{x}=f(x_{(t)},{\tilde{u}}_{(t)})$ is approximated using a fourth-order Runge-Kutta method.

    Although it would be ideal (especially in the fault case) to perform the dynamic programming algorithm in the 12-dimentional state space of the coupled rototranslational dynamics with twelve control variables, it would require an enormous amount of computing power and storage space for the quantized values. Therefore, a simpler solution is considered where $F_i^R$ and $M_i^B$ are computed separately for each channel as the optimal policies using the dynamic programming functional equation. Subsequently, using a quadratic programming control allocation method, a set of actuator commands $\bm{u}_c$ are found to generate forces and torques that match the optimal policies as closely as possible.

    To calculate $F_i^{\ast R}$ as the optimal policy, the following simplified state equations are used
    \begin{equation}
        \label{eq:xv}
    \begin{matrix}\left\{
        \begin{aligned}
    &{\delta \dot{x}}_i=\delta v_i\\
    &{\delta \dot{v}}_i=\frac{F_i^R}{m}\\
    \end{aligned}
    \right.&i=\left\{x,y,z\right\}\\
    \end{matrix}
    \end{equation}
    with quantized values of relative position and velocity as the states $\bm{x}=\left[\delta x_i \ \delta v_i\right]^T$ and $\tilde{u}=F_i^R$ as action. For rotational dynamics, we consider the following simplified formulation with states $\bm{x}=\left[\theta_i \ \omega_i\right]^T$ and $\tilde{u}=M_i^B$  to calculate the optimal torque $M_i^{\ast B}$ for channel $i$.
    \begin{equation}
        \label{eq:tw}
    \begin{matrix}\left\{
        \begin{aligned}
    &{\dot{\theta}}_i=\omega_i\\
    &{\dot{\omega}}_i=\frac{M_i^B}{J_i}\\
    \end{aligned}
    \right.&i=\left\{1,2,3\right\}\\
    \end{matrix}
    \end{equation}
    In the above equations, $m$ and $J_i$ represent the mass and principal moments of inertias, and the control bounds for the 12 thruster configuration shown in \figurename{~\ref{fig:SPHERESthrusters}} are
    \begin{equation}
        \label{eq:bounds}
        \begin{array}{c}
            - 2\sqrt{3}{u_{\max }} \le F_i^R \le 2\sqrt{3}{u_{\max }}\\
            - 2{u_{\max }}d \le M_i^B \le 2{u_{\max }}d 
           \end{array}
    \end{equation}
    The maximum force in the RSW frame ($F^R_i$) corresponds to when the net force of three pairs of thrusters firing along the x,y and z body axes aligns with an axis in the RSW frame. Additionally, the angle inputs to the attitude controllers $\left\{\theta_1,\theta_2,\theta_3\right\}$ are not calculated as a sequence of rotations with Euler angles. Instead, they are each considered as independent rotations about x, y, and z-axis of the coordinate system, respectively. Therefore, Eq.~\eqref{eq:q2angle} is used for this purpose to convert the quaternions to the controller input angles. This approach provides an overestimated value for the axis-angle rotation to the attitude controller and drives the error to zero faster than Euler angle rotations.
    \begin{equation}
        \label{eq:q2angle}
        \begin{matrix}
    \theta_i=2\,{\sin}^{-1}{(}q_i)&i=\left\{1,2,3\right\}
        \end{matrix}
    \end{equation}

    A control period of $\Delta t= 1$ second is considered as the time step for dynamic programming. It should be noted that the allotted thrusting duration for \textit{SPHERES} is only $200\,\si{\milli\second}$ in every 1 second, and the satellite spends the rest of the time updating its beacons \citep{Pong2010,Hilstad2010}. Because of this, the maximum effective thrust force in the control period is $u_\textrm{max}=0.2\,{u}_\textrm{on}$. Additionally, due to practical considerations, the on-off thrusters may take up to several milliseconds after receiving a command to change their state \citep{Hilstad2010}. Hence, a minimum pulse time of 10 milliseconds is considered as the time step for the simulations.
    
    In developing the performance measure of the dynamic programming problem (Eqs.~\eqref{eq:costfun} and \eqref{eq:gD}), the states are in fact the error between the actual states and their desired values, as in
    \begin{equation}
        \label{eq:transform_error}
        \bm{e}(t)=\bm{x}(t)-\bm{r}(t) 
    \end{equation}
    However, since no reference trajectory is assumed ($\bm{r}{(t)}=0$),  we are dealing with a regulation problem and the cost function reaches its minimum as the states and controls are driven to the origin. For attitude control, the attitude error should go to either zero or multiples of $2\pi$ in case of a flip. Otherwise, the satellite expends additional fuel on doing a reverse flip in order to reach $\theta_i=0$ where the cost is minimum. Therefore, in calculating the cost function for attitude dynamics, we define the attitude error as
    \begin{equation}
        \label{eq:attitude_error}
        \bm{e}_\theta=\sin{\left(\theta-\theta_r\right)} 
    \end{equation}
Finally, a general dynamic programming procedure in optimal control such as the one presented by \citet{Kirk2012} can be followed for computing the optimal policies. However, it can be seen from the dynamic programming functional equation \eqref{eq:func} that for a time-invariant system some terms always remain fixed for predefined quantized states and actions, and do not need to be re-calculated at every step of the computations. These values include the costs of going from one stage to the next, and the discrete next-stage states ($\bm{x}_{k+1}$) that result from applying predefined actions to states and are repeatedly called for interpolating values of $J^*$.  Instead, these values can be precomputed once and stored in memory to be recalled at every iteration. Therefore, the following algorithm is devised for computing the optimal policy that also utilizes multi-dimensional array operations for faster computations:

\begin{enumerate}
    \item {Create three $n_1\times n_2\times p$ rectangular grids $S_1$, $S_2$ and $U$ from the quantized arrays of states $\left\{s_1,s_2\right\}$ and action $\tilde{u}$ with lengths $n_1$, $n_2$ and $p$ respectively, such that
 \begin{equation}
    \label{eq:alg1}
\begin{cases}
S_1(i,j,k)=s_1(i)\\
S_2(i,j,k)=s_2(j)\\
U(i,j,k)=\tilde{u}(k)
\end{cases} \begin{matrix}
\textrm{for\ all} & 
\left( 
    \begin{matrix} 
i=\left\{1,\ldots,n_1\right\}\\
j=\left\{1,\ldots,n_2\right\}\\
k=\left\{1,\ldots,p\right\}\\
\end{matrix}\right)
\end{matrix} 
 \end{equation}
    }
    \item {Solve the systems of Eqs.~\eqref{eq:xv} and \eqref{eq:tw} with all admissible states and control values to find the resulting states after $\Delta t$, then store the values in two $n_1\times n_2\times p$ arrays $S_1^\prime$ and $S_2^\prime$ 
\begin{equation}
    \label{eq_xx2}
    \left[S_1^\prime,S_2^\prime\right]=f_D\left(S_1,S_2,U\right)
\end{equation}
    }
    \item{Calculate $J_\textrm{stage}$, a fixed $n_1\times n_2\times p$ matrix containing the costs for all combinations of states and actions. That is, $J_\textrm{stage}(i,j,k)$ stores the cost of taking action $\tilde{u}(k)$ with the initial states $s_1(i)$ and $s_2(j)$ from one stage to the next
    \begin{align}
        \label{eq:Jstg}
        & J_\textrm{stage}=Q_{s1} S_1^2+Q_{s2} S_2^2+R U \\
        & \bm{Q}=\left[\begin{matrix}Q_{s1}&0\\0&Q_{s2}\\\end{matrix}\right] \nonumber
    \end{align}
    where $Q_{s1}$, $Q_{s2}$, and $R$ are the cost weighting variables for $s_1$, $s_2$, and $\tilde{u}$, respectively.
    }
    \item{Initialize the $n_1\times n_2$ optimal value function at the final stage (terminal cost)
    \begin{equation}
        \label{eq:termcost}
    J^\ast(i,j)=\left[\begin{matrix}s_1(i)&s_2(j)\\\end{matrix}\right]\tilde{\bm{H}}\left[\begin{matrix}s_1(i)&s_2(j)\\\end{matrix}\right]^T 
    \end{equation}
    }
    where $\tilde{\bm{H}}$ is the weight matrix related to the terminal cost.
    \item{Update the optimal value function by iterating over the following functional equation $N$ times
    \begin{equation}
        \label{eq:iterfun}
        J_\textrm{next}^\ast = \underset{\tilde{u}}{\text{ min}} \left\{
            J_\textrm{stage}+J_\textrm{prev}^\ast(S_1^\prime,S_2^\prime)
        \right\} 
    \end{equation}}
\end{enumerate}
The first term is a fixed 3-dimensional array containing the cost for all states and actions taken for one stage. The second term contains the cost of the optimal k-stage trajectory towards the terminal state. The minimization returns the smallest values along the ${3^\textrm{rd}}$ dimension (where $\tilde{u}$ varies) of the multidimensional array, updating the optimal value function. Finally, the indices of the minimum values along the ${3^\textrm{rd}}$ dimension are stored as an $n_1\times n_2$ matrix $I^*$, and the optimal policy is found by calculating $U^*\left(i,j\right)=\tilde{u}\left(I^*(i,j)\right)$ for all elements in $I^*$.

\subsection{Control Allocation and Pulse Modulation} \label{sec:control_aloc}

There are several control allocation methods that can be used to compute a set of thruster on-off commands that produce the desired controls. A detailed evaluation of some of the optimization-based algorithms for control allocation is provided by \citet{Bodson2002}. For thruster mapping, simplex-based methods \citep{Yang2003}, redistributed pseudoinverse \citep{Pong2010}, and quadratic-programming-based control allocation \citep{Pong2010} have been used. The pseudoinverse method is simple but is not guaranteed to always find the actuator commands that produce the closest required controls. On the other hand, the quadratic programming problem is formulated as finding a set of thruster commands ($\bm{u}_\textrm{control}$) so that the error between the generated controls and the desired controls is minimized
\begin{equation}
    \label{eq:quadprog}
    \begin{aligned}
         { \bm{u}_\textrm{control}} = & ~ \underset{\tilde{u}_{N-K}}{\text{ { arg\,min}}} \left\{
        \frac{1}{2}\left(\bm{Hu}-\left[\begin{matrix}\bm{F}^{*B}&\bm{M}^{*B}\\\end{matrix}\right]^T\right)^T\bm{W}\left(\bm{Hu}-\left[\begin{matrix}\bm{F}^{*B}&\bm{M}^{*B}\\\end{matrix}\right]^T\right)\right\}\\
     & \text{subject to} \quad u_j\in U_\textrm{feasible} 
    \end{aligned}
    \end{equation}
    This optimization can be performed separately for each channel described by Eq.~\eqref{eq:channel} with $\bm{H}_c$ that indicates the arrangement of the four thrusters in each channel, and
    \begin{equation}
        \label{eq_xx1}
    \bm{W}=\left[\begin{matrix}W_f&0\\0&W_m\\\end{matrix}\right]
    \end{equation}
    to weigh the relative importance of translational and attitude control. To address the different ranges of values for $\bm{F}^*$ and $\bm{M}^*$ and to deal with only one parameter, the weighting matrix can be modified as follows
    \begin{align}
        \label{eq:Wmf}
    & \bm{W}^\prime=\left[\begin{matrix}1&0\\0&W_{mf}\\\end{matrix}\right] \\
    & W_{mf}=\frac{W_m}{d \ W_f}  \nonumber
    \end{align}
    Hence, a value of $W_{mf}=1$ in this formulation corresponds to assigning equal weights on attitude and position control, whereas a slightly higher/lower value of $W_{mf}$ puts more priority on attitude/position control, respectively. 

\begin{figure}[!htp]
    \centering
    {\includegraphics[width=0.68\columnwidth]{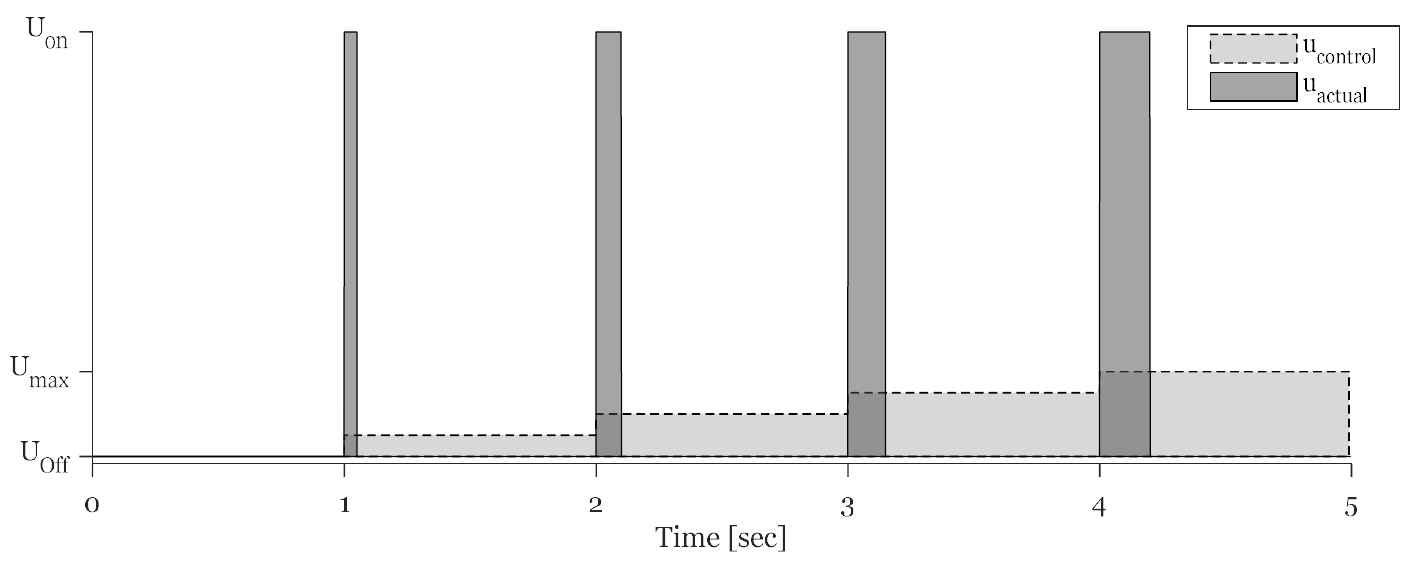}}
    \caption{Pulse modulation of $\bm{u}_\textrm{control}$ signals to actual thruster commands}
    \label{fig:pwm}
  \end{figure}

    To solve the quadratic programming problem, different algorithms for finding the best command strategy can be used \citep{Pong2010}. However, the actuator commands within the bounds described by Eq.~\eqref{eq:bounds} can be quantized into values that the thrusters can effectively produce in increments of 10 milliseconds (up to 200~\si{\milli\second} in the control period). Since the actuator commands are predetermined values, the minimum of Eq.~\eqref{eq:quadprog} can be found by putting in all possible combinations of thruster on-offs in the control period, excluding the cases where back-to-back thrusters are both ``on'' and their forces cancel out. For each pair of thrusters, there are collectively 20 positive forces, 20 negative forces, and 1 case of $\bm{u}_j = 0$. The resulting set of all admissible cases is denoted by $U_\textrm{feasible}$. The computing time for this approach is quite low as the search space for one channel only comprises 1681 ($ = 41 \times 41$) cases. When one thruster has failed, the search space for the corresponding channel further reduces to 861 ($= 41 \times 21$) cases. During simulations, the effective $\bm{u}_\textrm{control}$ from the optimization is converted to actual thruster-on times for one control period using a pulse modulation technique, i.e., by conserving the impulse of the modulated signal. Figure \ref{fig:pwm} illustrates a sample of the resulting actuator commands using pulse modulation. Thruster-on times, in milliseconds, are determined by the following relation for each control period:
    
\begin{equation}
    \label{eq:t_on}
\bm{t}_\textrm{on}=200 \ \frac{\bm{u}_\textrm{control}}{u_\textrm{max}}
\end{equation}

\subsection{Control in Thruster Failure Mode} \label{sec:failure}

As mentioned in Section~\ref{sec:dynamic_prog}, if the practical limitations on storage space and computational power did not exist, the optimal actuator commands could be computed directly as dynamic programming policies. Nevertheless, the proposed dynamic-programming-based method utilizes a control allocator to map the commands to thruster firings, so this module must be reconfigured when one or more thrusters fail. The feasibility of using reconfigurable control allocation has been previously studied by \citet{Pong2010}. It was demonstrated that PD control along with control allocation works when the satellite approaches the target from the side with two operative thrusters. In this case, the satellite has full rotational control while simultaneously driving its linear velocity to zero. However, when the satellite approaches the target from the side with only one operative thruster, any attempt to stop the spacecraft using that single thruster would cause a spinning motion about another direction. For \textit{SPHERES}, the angular acceleration generated by one thruster is so high compared to its linear acceleration that the satellite would be left spinning in place if it uses only one thruster to stop. Therefore, the control allocator must handle the dynamic coupling and keep the underactuated spacecraft stable while following the commands from dynamic programming controllers as closely as possible.

In this paper, it is assumed that the faulty thruster has already been identified by a fault detection and isolation (FDI) system. The proposed control in thruster-failure mode is then developed in three steps. First, the control bounds of Eq.~\eqref{eq:bounds} are updated to include only the maximum forces and torques that can be produced by the operative thrusters, and dynamic programming optimal policies are computed again. Next, the admissible set of actuator commands in Eq.~\eqref{eq:quadprog} is recalculated after excluding the failed thruster. Finally, the weighting variable $W_{mf}$ is dynamically adjusted depending on the relative position to ensure a stable control when the spacecraft approaches the target from the side with a failed thruster, or force the spacecraft to approach the target from the side with two operative thrusters. Whenever the initial position of the spacecraft is on the desirable side, a value of $0.9\leq W_{mf} \leq 1.1$ would be sufficient for a stable control. In fact, it was found by trial and error that a value of $W_{mf}=0.93$ shows the best performance and the lowest fuel consumption. This way, the control allocator puts more priority over position control and consequently does not allow the spacecraft to overshoot the target position. Furthermore, since the spacecraft is underactuated, there is a stability constraint on the thruster commands from the quadratic programming formulation in Eq.~{\ref{eq:quadprog}}. Therefore, whenever the spacecraft is approaching the target from the unstable side, a higher value is set for $W_{mf}$ (i.e., more weight on attitude control) so that the spacecraft will temporarily suspend position control if there is no command strategy for controlling the position without disturbing the attitude stability. This will cause the spacecraft to intentionally miss the target position to reach an arbitrary point beyond it, where the control allocation weight is switched to the constant described in the former case which will cause the spacecraft to perform a maneuver back to the actual target.

In order to reduce the overshoot and fuel consumption of the maneuvers, the weighting variable is tuned for different initial positions. Starting from relatively close distances, a value for $W_{mf}$ is found for each maneuver to achieve a stable control and to minimize the overshoot. This process is continued for distances up to 10~\si{\meter} and, at every iteration, the weighting variable is linearly interpolated between the values previously obtained. This tuning process converges into a stable control for all distances in the range. Lastly, the process is repeated for each of the 12 cases of thruster failures. 

\subsection{Lyapunov-Based Thruster Selection Method} \label{sec:lyap}

The Lyapunov-based thruster selection algorithm \citep{Curti2010} is used for comparing the effectiveness of the proposed dynamic-programming-based method. While only a brief explanation of the approach is offered in this section for ease of understanding the results of the analysis, the reader is encouraged to consult the reference \citep{Curti2010} for full details on the activation algorithm, the auxiliary variables, and the dynamics representation with modified Rodriguez Parameters (MRPs). 

The Lyapunov-based algorithm works by defining a set of auxiliary switching variables based on the tracking error between the satellite actual states and the decoupled reference models described by Eq.~\eqref{eq:refmodels}

\begin{equation} 
    \label{eq:refmodels}
    \begin{cases}
    &\ddot{\bm{\rho}}_m+K_1\dot{\bm{{\rho}}}_m+K_2\bm{\rho}_m=\bm{v}_{\rho c}\\
    &\ddot{\bm{\sigma}}_m+K_3\dot{\bm{\sigma}}_m+K_4\bm{\sigma}_m=\bm{v}_{\sigma c}
    \end{cases}
\end{equation}
where $\bm{\rho}_m$ is the relative position vector of the reference trajectory, the MRPs are expressed in terms of the quaternions as $\bm{\sigma}_m^T=\bm{q}_{13}/q_4$, and $\bm{v}_{\rho c}$ and $\bm{v}_{\sigma c}$ are the reference inputs.

The Lyapunov-based approach guarantees asymptotic stability of the tracking error dynamics as long as the parameters $\left\{K_1,\ldots,K_4\right\}$ of the reference models are designed in a way that the defined ideal controls $\bm{\bar{w}}_F$ and $\bm{\bar{w}}_M$ which drive the tracking error exponentially to zero, never exceed the maximum thrusting capability of the satellite. 
For coupled rototranslational control of \textit{SPHERES}, the theoretical bounds on the ideal controls are
\begin{equation}
    \label{eq:lyap_constr}
    \begin{cases}
        \left|w_{Fi}\right|\le u_\textrm{on}\\
        \left|w_{Mi}\right|\le d \ u_\textrm{on}
    \end{cases} i=\left\{1,2,3\right\} 
\end{equation}
The thruster selection algorithm then determines a thrusting strategy with minimum active thrusters that drives the tracking error to zero.
Moreover, the parameters $\left\{K_1,\ldots,K_4\right\}$ are tuned based on the initial conditions of the maneuver to ensure the limits on $\bm{\bar{w}}_F$ and $\bm{\bar{w}}_M$ are never breached. As a result, a set of parameters tuned for a short distance maneuver may no longer work for long re-locations. This issue is demonstrated in a sensitivity analysis in Section~\ref{subsec:sim_operative}. 

Thrusting is performed with an update rate of 10 milliseconds that is the minimum pulse time of the thrusters. To account for the \textit{SPHERES}' thrusting schedule, Lyapunov-based thrusting is activated only in the first 200~\si{\milli\second} in every 1 second. This will result in a slower-moving response as it reduces the effective thrust force, and the satellite's attitude would drift during the 800~\si{\milli\second} downtime as the Lyapunov-based method does not incorporate any predictive mechanism. Despite changing the accuracy, this policy does not affect the fuel consumption ratios obtained in Section~\ref{subsec:sim_operative} for the comparison because the reference models are still fitted to the dynamic programming responses. This has been verified through numerical simulations conducted with and without this policy. Similarly, increasing the minimum pulse time of the thrusters does not result in any noticeable improvement in total fuel consumption, but causes a notable loss in accuracy. 

The Lyapunov thruster selection approach cannot be used for comparison in the fault case as the method only addresses control for a fully actuated spacecraft with a minimum of 12 thrusters. In case one or more thrusters fail, the spacecraft cannot control its position without altering its attitude and vice versa. Being capable of performing translational-only and rotational-only maneuvers is a requirement that the Lyapunov thruster selection approach is based on. \citep{Curti2010}

\section{Simulation Results} \label{sec:simul}

\begin{figure}[!htp]
    \centering
    \subfloat[]{ {\includegraphics[width=0.8\textwidth]{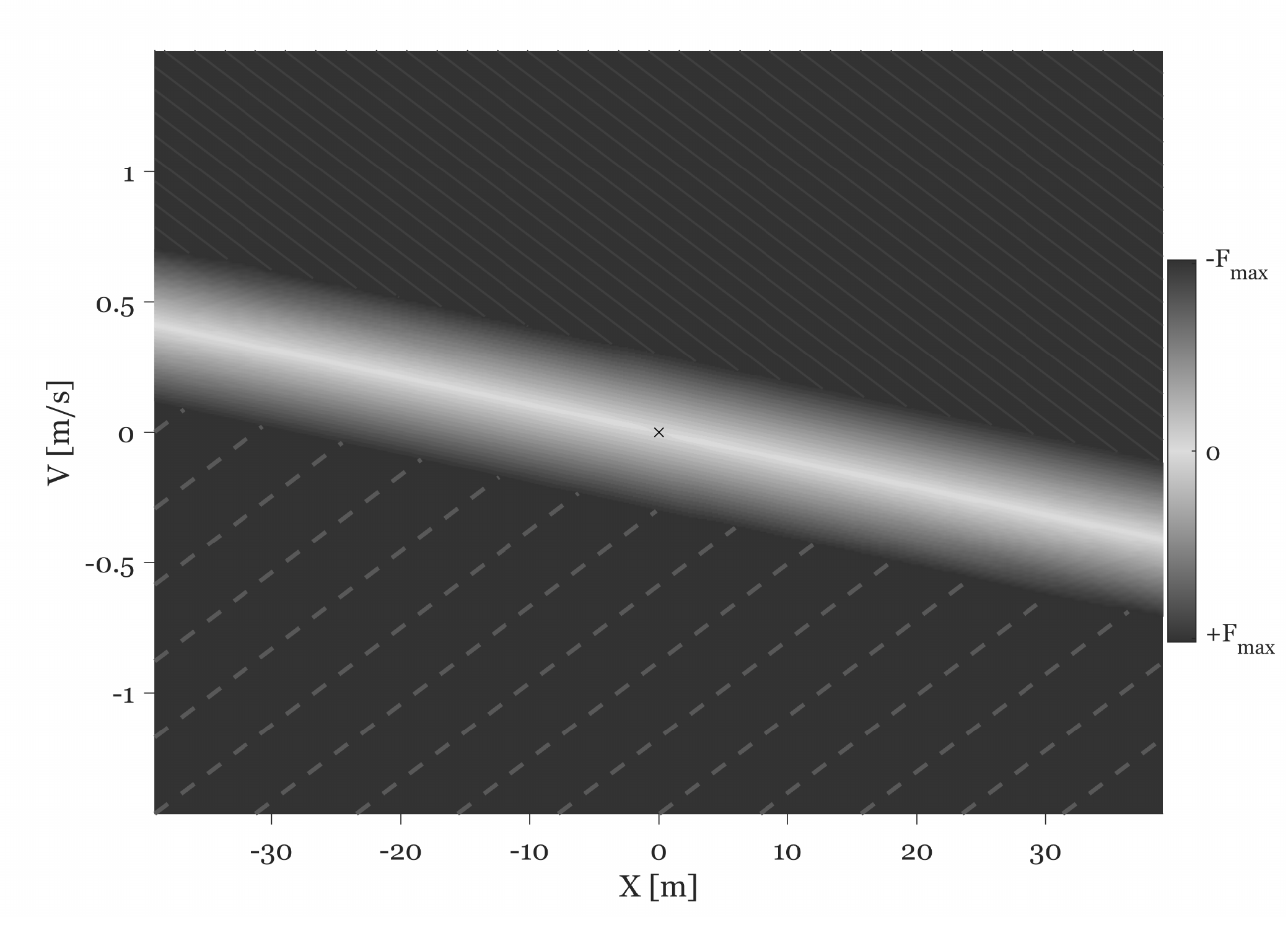}} }%
    \qquad
    \subfloat[]{ {\includegraphics[width=0.8\textwidth]{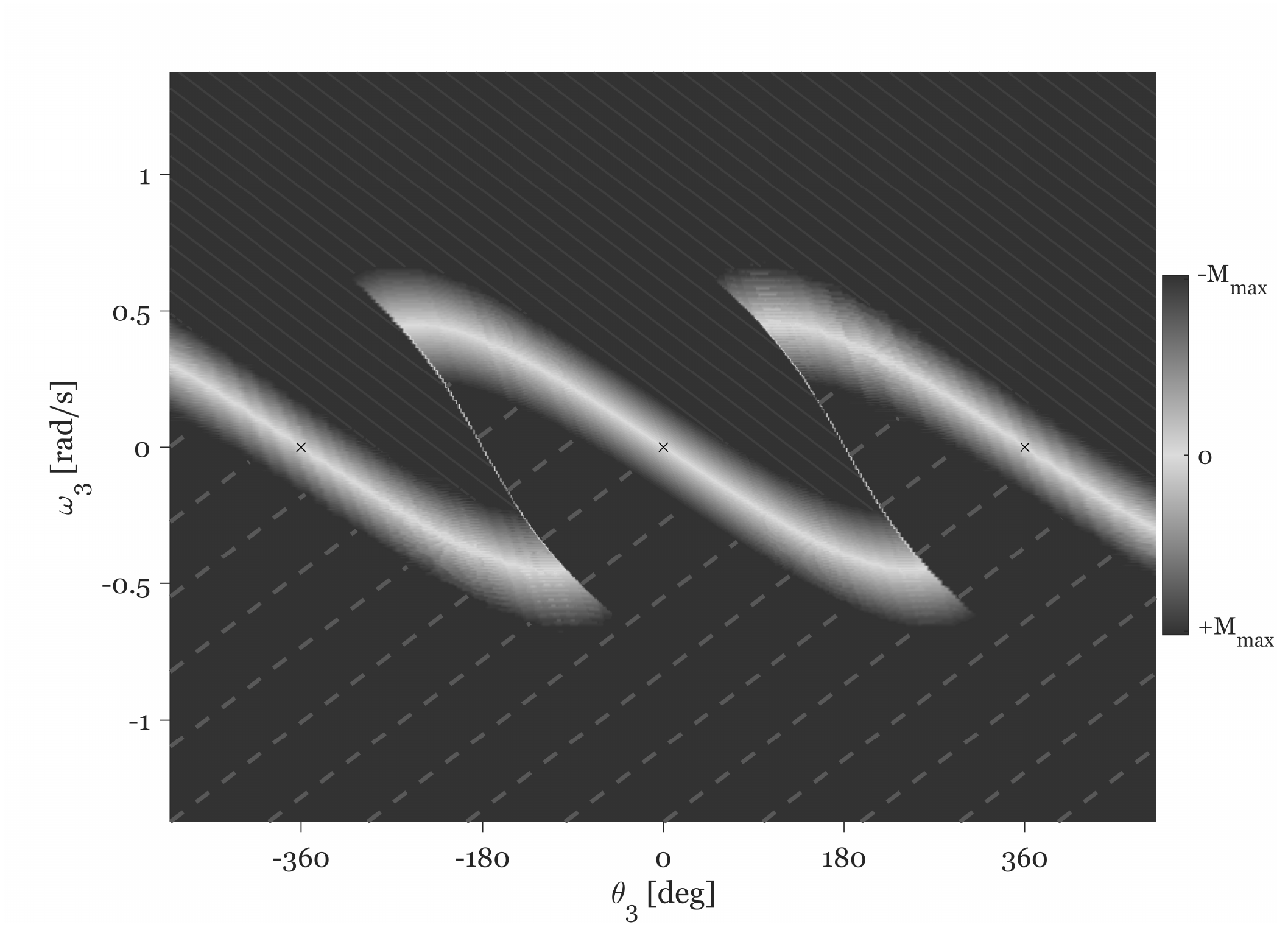}}}%
    \caption{Dynamic programming optimal policies for (a) position control and (b) attitude control}
    \label{fig:dppolicy}
\end{figure}

\begin{figure}[!htp]
    \centering
    {\includegraphics[ width=0.7\columnwidth]{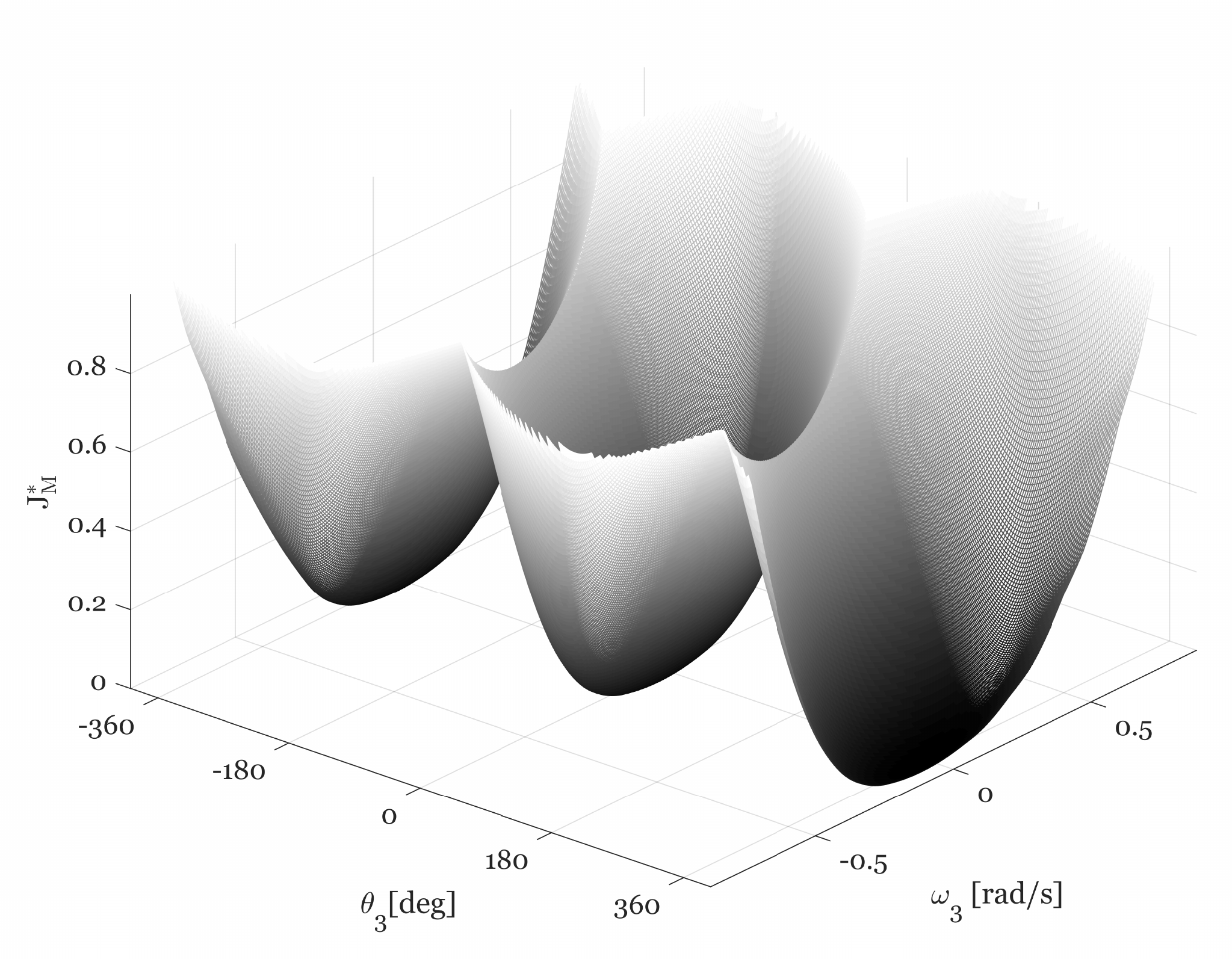}}
    \caption{Optimal cost $J^\ast$ for moment control}
    \label{fig:JM}
\end{figure}

This section presents the results of 6-DOF simulations to demonstrate the performance of the proposed control method in two cases of thruster failures and to compare it with the Lyapunov thrusters' selection approach when all thrusters are operative. First, the optimal control policies are obtained by running dynamic programming using the parameters listed in Table~\ref{tab:DPparams}. The resulting policies for the force and torque controllers are shown in \figurename{~\ref{fig:dppolicy}}, where the points in the area shaded with straight lines correspond to the maximum negative values of forces or torques, and the points shaded with dashed lines correspond to the maximum positive forces or torques. The white line that crosses the origin represents the dead zone - where the controls are zero - and the in-between values gradually become darker as they approach the maximum values. The width of the dead zone is proportional to the ratio $\frac{Q_{s1}}{R}$, a trade-off between fuel expenditure and the steady state error, whereas the ratio $\frac{Q_{s2}}{Q_{s1}}$ changes the slope of the line and is used for adjusting the damping of the response. These attributes of the dynamic-programming-based approach can help in determining the parameters of the quadratic cost function in order to reach desired steady-state errors. For instance, the parameters in Table~\ref{tab:dpfault} are designed such that the maximum width of the dead-zones for position and attitude are less than 5 \si{\milli\meter} and 0.6 degrees, respectively. Furthermore, as a result of implementing Eq.~\eqref{eq:attitude_error}, the optimal cost for attitude control reaches zero at multiples of $2\pi$ as can be seen in the example shown in \figurename~\ref{fig:JM}.

{\begin{table}[htbp]
    \centering
    \caption{Dynamic programming parameters for the fully-actuated scenario}
    \begin{tabular}{lcc}
        \toprule
        \multirow{2}{*}{Parameters} & \multicolumn{2}{c}{Value}\\
        \cmidrule{2-3}
        & Position Control & Attitude Control \\
        \midrule 
        \multirow{2}{*}{
            \begin{minipage}[t]{0.2\linewidth}%
                Selected range
                 of state variables %
            \end{minipage} 
        }& $-40\,\si{\meter}\le x\le40\,\si{\meter}$ & $-400\si{\degree}\le \theta \le 400\si{\degree}$ \\
        & $-1.5\,\si{m/s}\le v \le1.5\,\si{m/s}$ & $-2.6\,\si{\radian/\second}\le \omega \le2.6\,\si{\radian/\second}$ \\
        $Q$  & $10^{-4}\left[\begin{matrix}0.03&0\\0&21\\\end{matrix}\right]$ & $\left[\begin{matrix}0.1&0\\0&0.5\\\end{matrix}\right]$ \\
        $R$ & 1 & 1\\
        $T_{\textrm{final}}$ & 500\,\si{s} & 200\,\si{s}\\
        Time step ($\Delta t$) & 1\,\si{s} & 1\,\si{s} \\
        \bottomrule
    \end{tabular}
    \label{tab:DPparams}
\end{table}}

For each of the state variables, 700 logarithmically spaced points are generated within the specified ranges. Logarithmic spacing effectively increases the resolution around the origin - where it is most desired - by creating a denser meshing, while eliminating unnecessary computations in peripheral regions. For the control variables, 41 linearly spaced points are generated to include all the feasible actions that the thrusters are capable of producing. Furthermore, the terminal cost (first term of Eq.~\eqref{eq:costfun}) is omitted as a sufficiently high $\bm{Q}$ in the second term of Eq.~\eqref{eq:costfun} would automatically drive the states to zero. Also, if a predefined terminal state or a reference trajectory is desired for a particular problem, the optimal policies can be found by transforming the states using Eq.~\eqref{eq:transform_error}.

\subsection{Simulation with All Thrusters Operative} \label{subsec:sim_operative}

\begin{figure}[!htp]
    \centering
    {\includegraphics[ width=0.95\columnwidth]{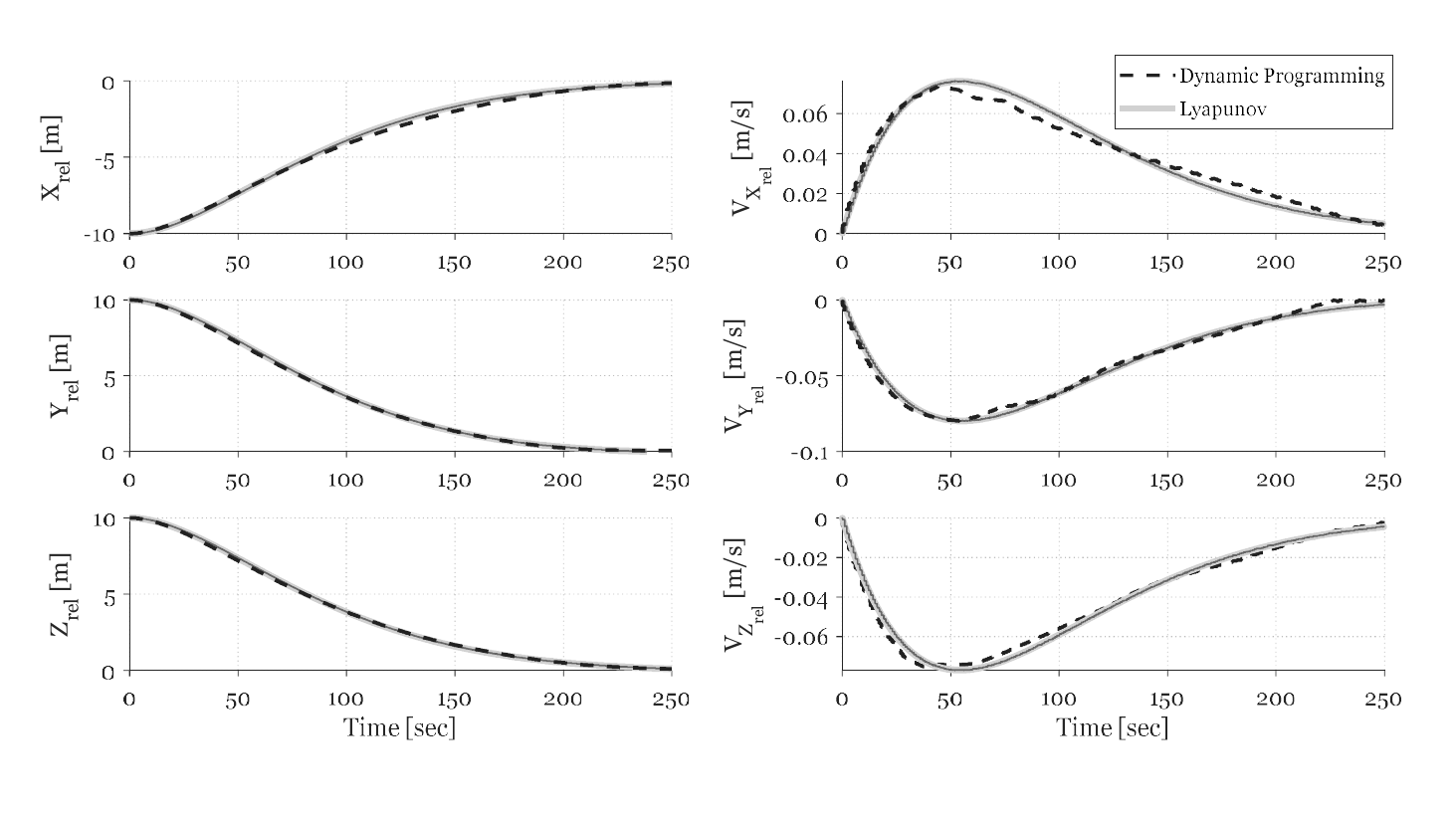}}
    \caption{Positions and velocities in the fully-actuated scenario}
    \label{fig:comparison_pv}
  \end{figure}

\begin{figure}[!htp]
    \centering
    {\includegraphics[width=0.85\columnwidth]{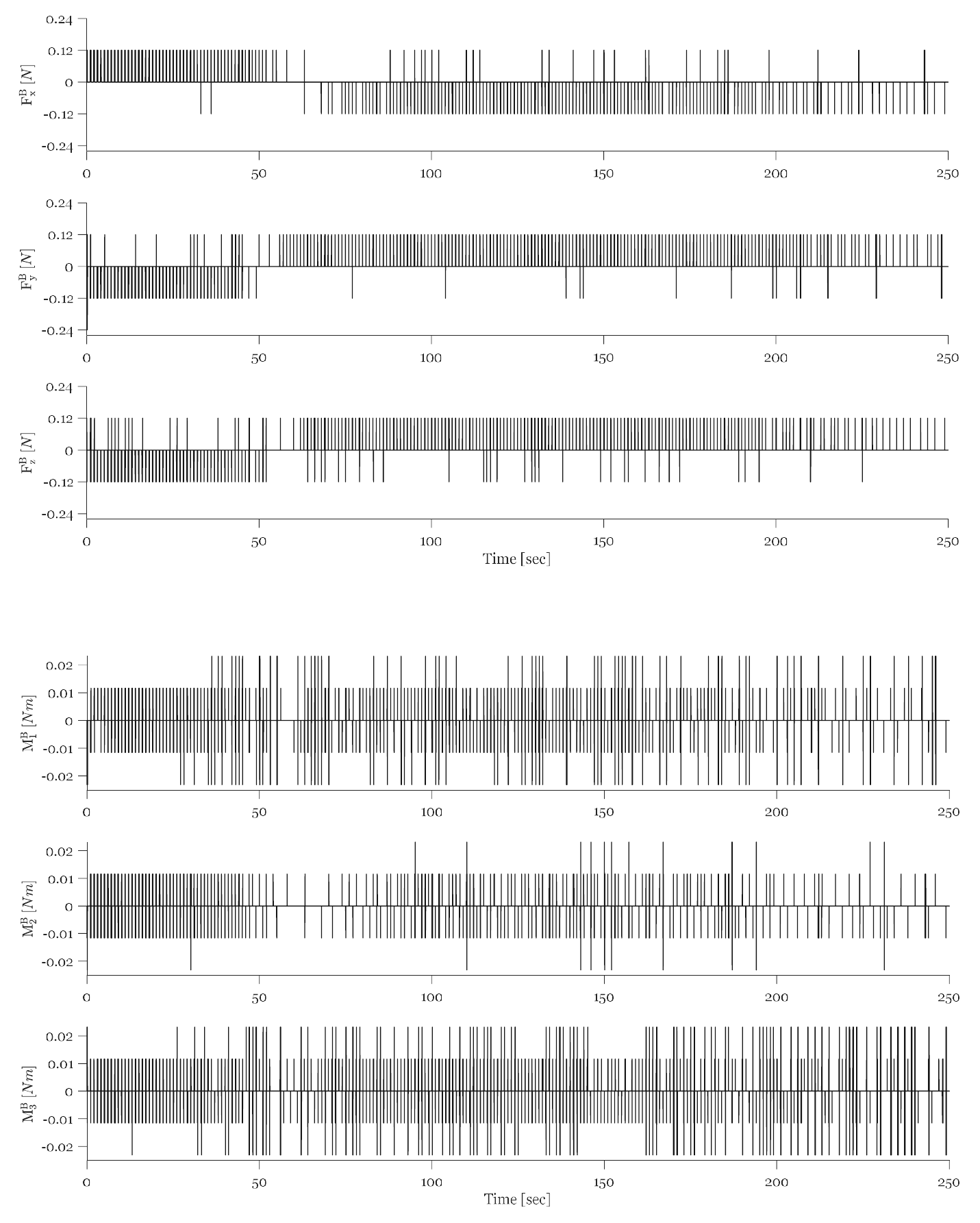}}
    \caption{Thruster-generated body forces and torques for Lyapunov-based control in the fully-actuated scenario}
    \label{fig:comparison_FM_lyap}
  \end{figure}

The results of the 6-DOF simulations for the proposed dynamic programming control along with the Lyapunov-based control are presented in \figurename{~\ref{fig:comparison_pv}} to \figurename{~\ref{fig:comparison_FM_lyap}}. The fully-actuated satellite starts from the initial conditions $\bm{\rho}_0=\left[-10 \ \ 10 \ \ 10\right]^T$ and $\bm{\theta}_0=\left[30 \si{\degree} \ \ 30 \si{\degree} \ -30 \si{\degree}\right]^T$ with zero initial velocities, and moves to zero out its position and attitude. The state vector of the satellite is $\bm{x}_{sat} = \left[ \bm{\rho}, \ \dot{\bm{\rho}}, \ \bm{q}, \ \bm{\omega} \right]^T$. For control allocation, the value for $W_{mf}$ is set to unity. For Lyapunov-based control, $\bm{v}_{\rho c}=0$ and $\bm{v}_{\sigma c}=0$ are the default values for the regulation problem \citep{Curti2010}. The orbital parameters of the International Space Station (ISS) are also used in simulating the orbit.

  \begin{figure}[!htp]
    \centering
    {\includegraphics[width=0.95\columnwidth]{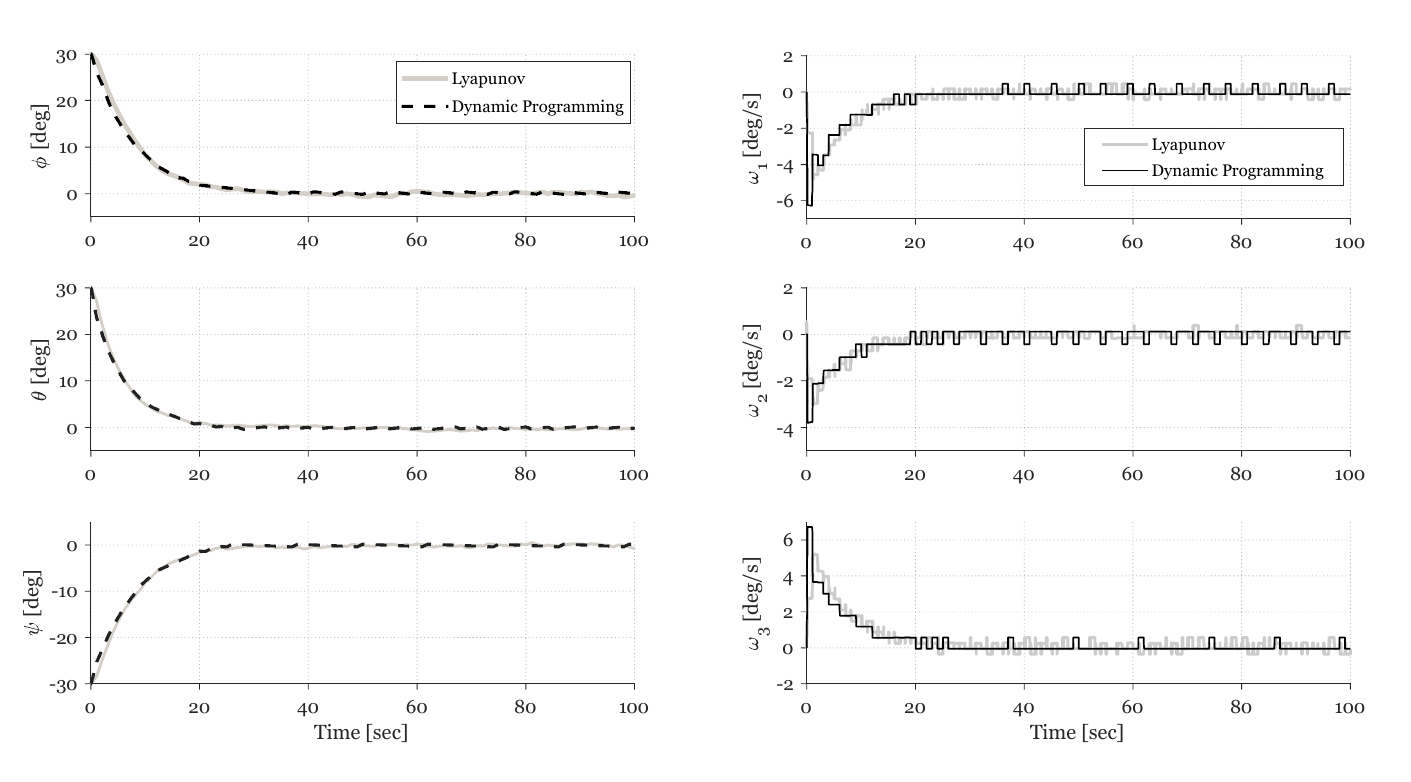}}
    \caption{Angles and rotational velocities in the fully-actuated scenario}
    \label{fig:comparison_tw}
  \end{figure}

  \begin{figure}[!htp]
    \centering
    {\includegraphics[width=0.85\columnwidth]{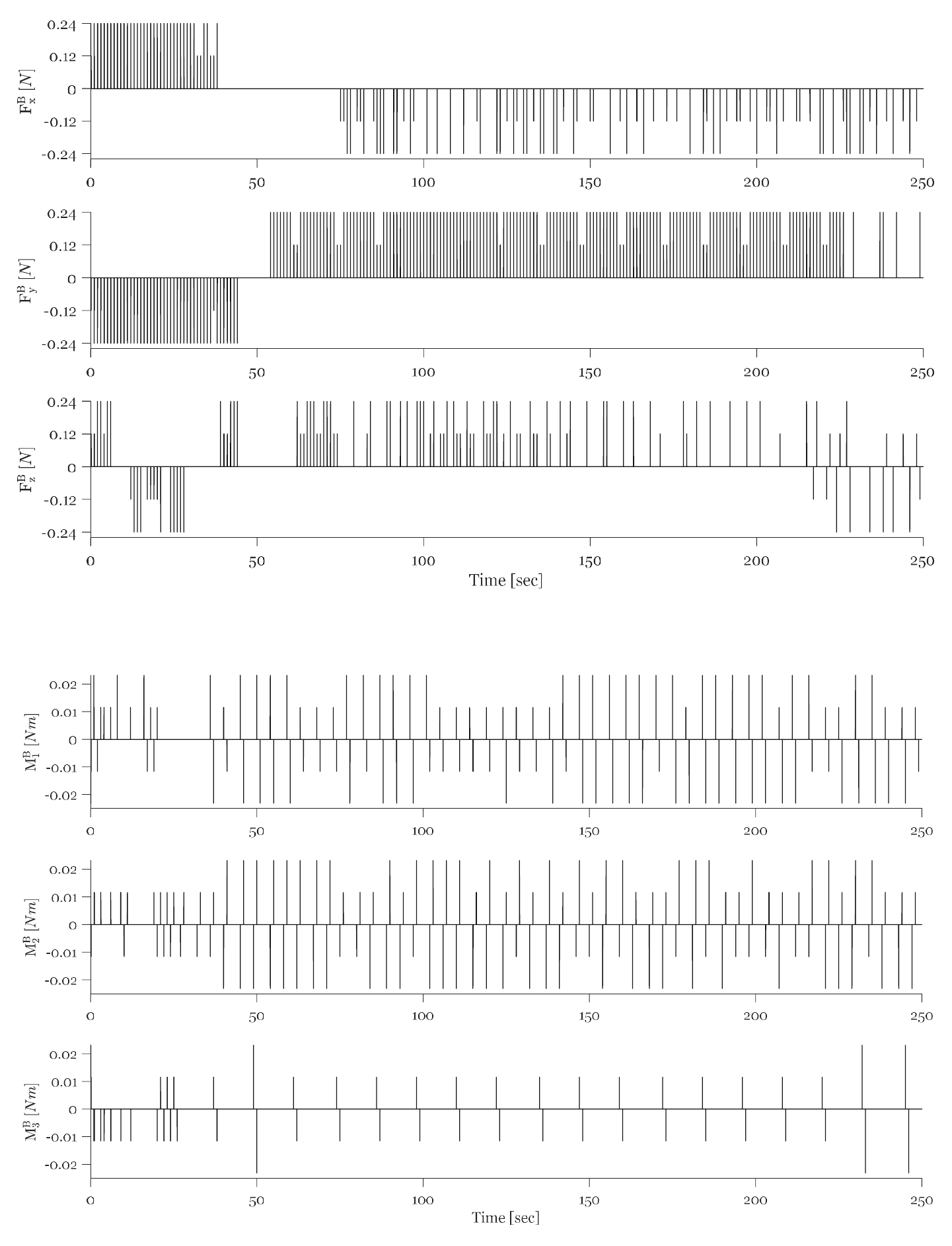}}
    \caption{Thruster-generated body forces and torques for dynamic programming control in the fully-actuated scenario}
    \label{fig:comparison_FM_dp}
  \end{figure}

To use the fuel consumption of maneuvers as a valid criterion for comparison, the reference models of the Lyapunov-based control are optimized so that both control systems have similar responses with matching settling times~($t_s$). This is achieved by using a generalized pattern search algorithm \citep{Audet2002} to find the best set of parameters $\left\{K_1,\ldots,K_4\right\}$ that minimize the position and attitude trajectory errors between the Lyapunov approach and the dynamic programming control. Table~\ref{tab:DPparams} and Table~\ref{tab:Lyapparams} summarize the parameters of the two methods.

{\setlength{\aboverulesep}{0pt}
\setlength{\belowrulesep}{0pt}
\begin{table}[htbp]
    \centering
    \caption{Tuned parameters for the Lyapunov-based control}
    \begin{tabular}{cc|cc}
        \toprule
        \multicolumn{2}{c}{Position Control} & \multicolumn{2}{c}{Attitude Control}\\
        \cmidrule{1-2} \cmidrule{3-4}
        $K_1$ & $10^{-2} \textrm{diag}\left\{{3.44,\ 3.20,\ 3.39}\right\}$
        & $K_3$ & $\textrm{diag}\left\{{3.31,\ 4.60,\ 3.23}\right\}$ \\
        $K_2$ & $3.7\times10^{-4} \bm{I}_{3\times3}$
        & $K_4$ & $\textrm{diag}\left\{{0.47,\ 0.69,\ 0.48}\right\}$ \\
        $Q_{1L}$ & $5\times10^4 \bm{I}_{3\times3}$
        & $Q_{2L}$ & $5\times10^5 \bm{I}_{3\times3}$ \\
        \bottomrule
    \end{tabular}
    \label{tab:Lyapparams}
\end{table}}

Figures \ref{fig:comparison_pv} and \ref{fig:comparison_tw} show the matched position and attitude responses. After the settling time, the states begin to jitter around the desired points because the on-off thrusters have a minimum pulsing time and are incapable of making the error precisely zero. As can be seen in \figurename~\ref{fig:comparison_FM_lyap}, thrusters' commanding in the Lyapunov-based control is quite excessive compared to that of the dynamic programming control (\figurename~\ref{fig:comparison_FM_dp}). It is also observed that the Lyapunov-based method only activates one thruster along each force component in controlling the position, which asserts that the activation algorithm never switches to position-only control. This is because in \textit{SPHERES}'s case even 10 \si{\milli\second} thruster reactions produce angular accelerations high enough to cause the attitude to go back and forth across the desired angle. Therefore, attitude control is always in action. Additional details of the simulations are provided in Table~\ref{tab:Simresults}. Based on these details, the response of the Lyapunov-based method exhibits more steady-state errors in position and attitude, and more than double fuel consumption in terms of total impulse.

{\begin{table}[htbp]
    \centering
    \caption{Numerical results of the dynamic-programming and Lyapunov-based methods}
    \begin{tabular}{lcc}
        \toprule
        Parameter & Dynamic Programming Control & Lyapunov-based Control\\
        \midrule
        Settling Time  & 238.5\,\si{\second} & 243.7\,\si{\second} \\
        Total Impulse  & 2.03\,\si[inter-unit-product = {.}]{\newton\second} & 4.62\,\si[inter-unit-product = {.}]{\newton\second} \\
        Position Max Steady-State Error  & 14.01\,\si{\centi \meter} & 16.67\,\si{\centi \meter} \\
        Attitude Max Steady-State Error & 0.52\si{\degree} & 1.06\si{\degree} \\
        \bottomrule
    \end{tabular}
    \label{tab:Simresults}
\end{table}}

\begin{figure}[!htp]
    \centering
    {\includegraphics[width=0.65\columnwidth]{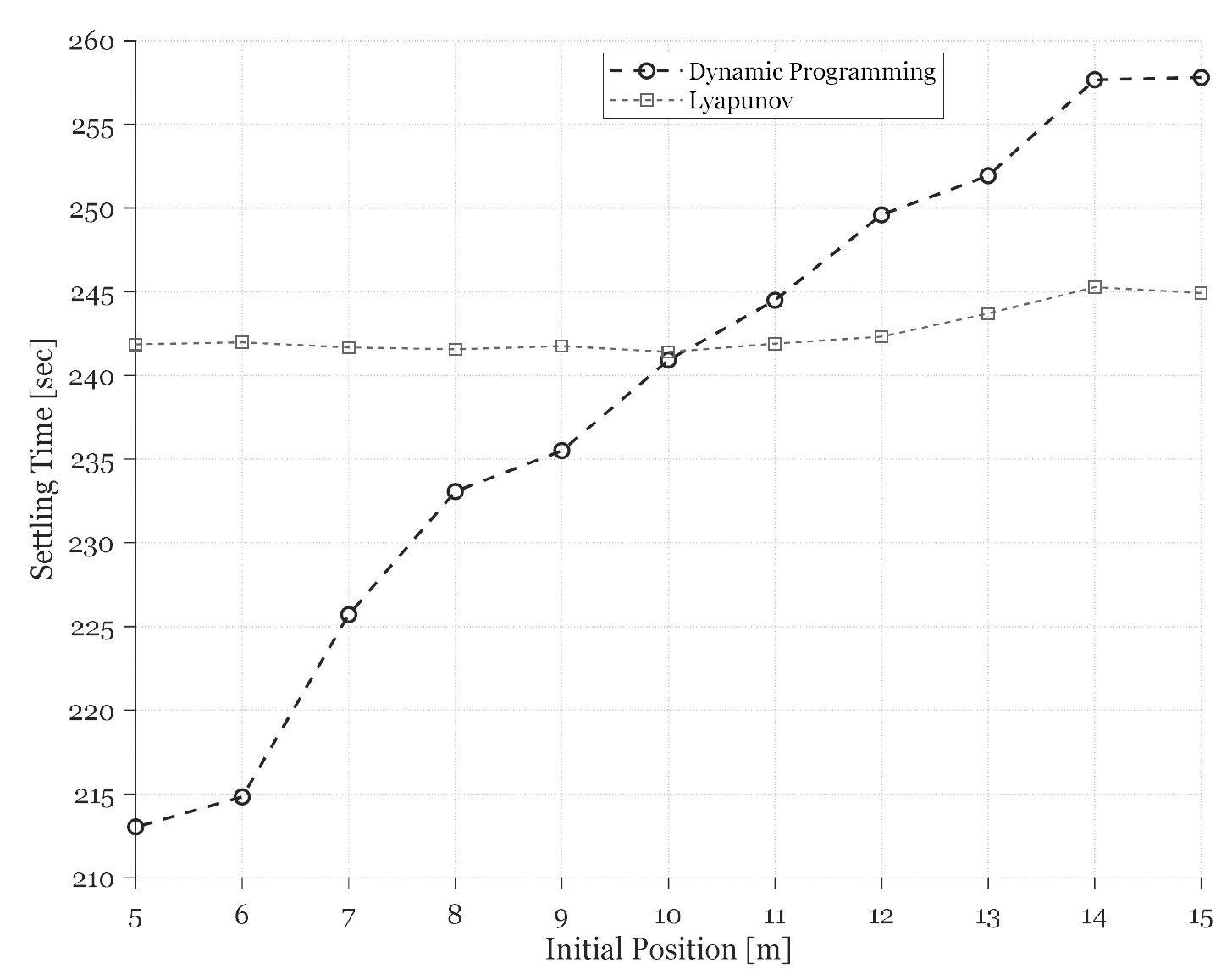}}
    \caption{Max settling times as a function of the initial position $\bm{\rho}_{0}=\left[r \ \ r \ \ r\right]^T$}
    \label{fig:maxsettle}
  \end{figure}

\begin{figure}[!htp]
    \centering
    {\includegraphics[width=0.65\columnwidth]{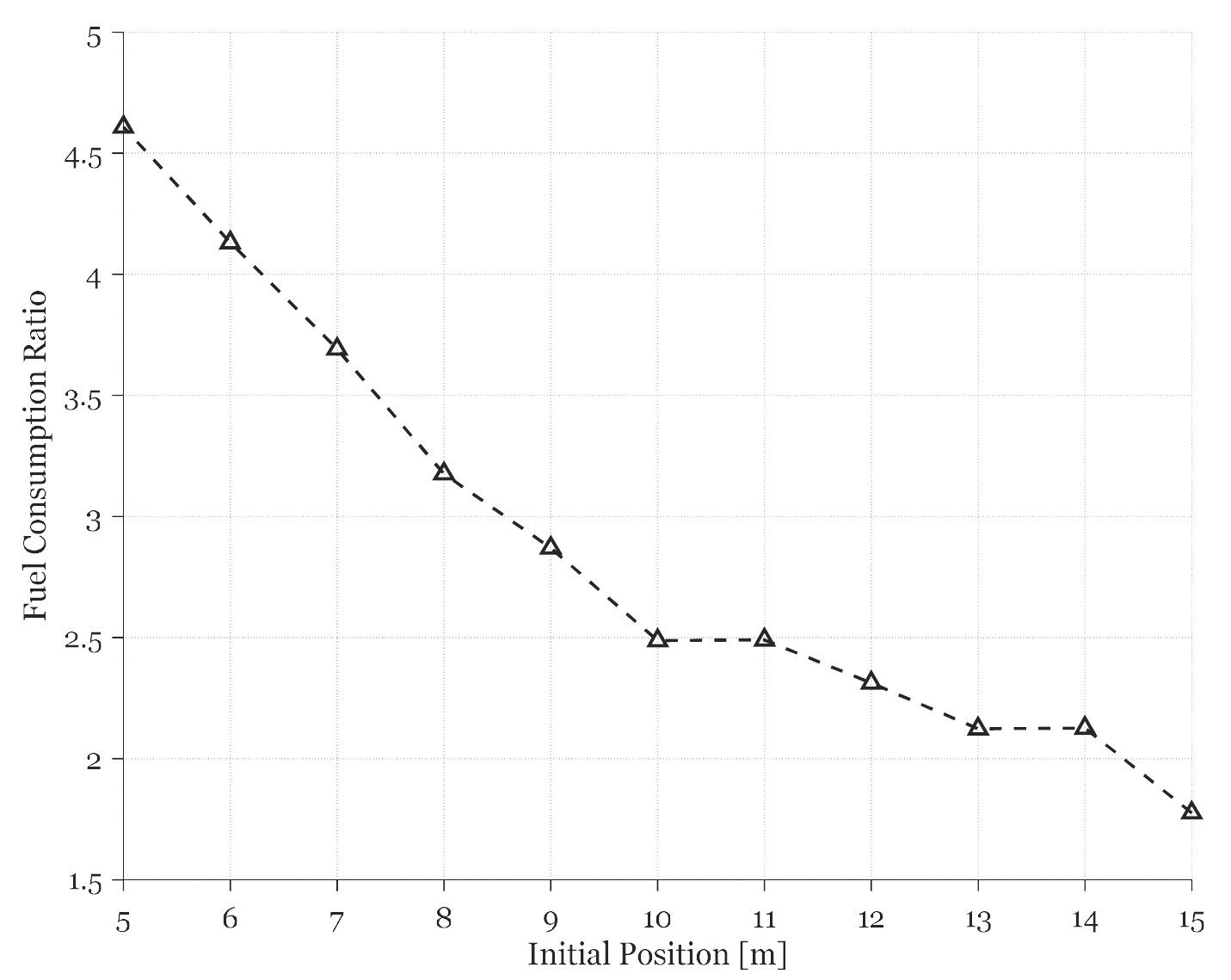}}
    \caption{Ratio of impulses (Lyapunov-based control to dynamic programming) as a function of the initial position $\bm{\rho}_{0}=\left[r \ \ r \ \ r\right]^T$}
    \label{fig:impulses}
  \end{figure}

\begin{figure}[!htb]
    \centering
    \minipage{0.32\textwidth}
    {\includegraphics[width=\linewidth]{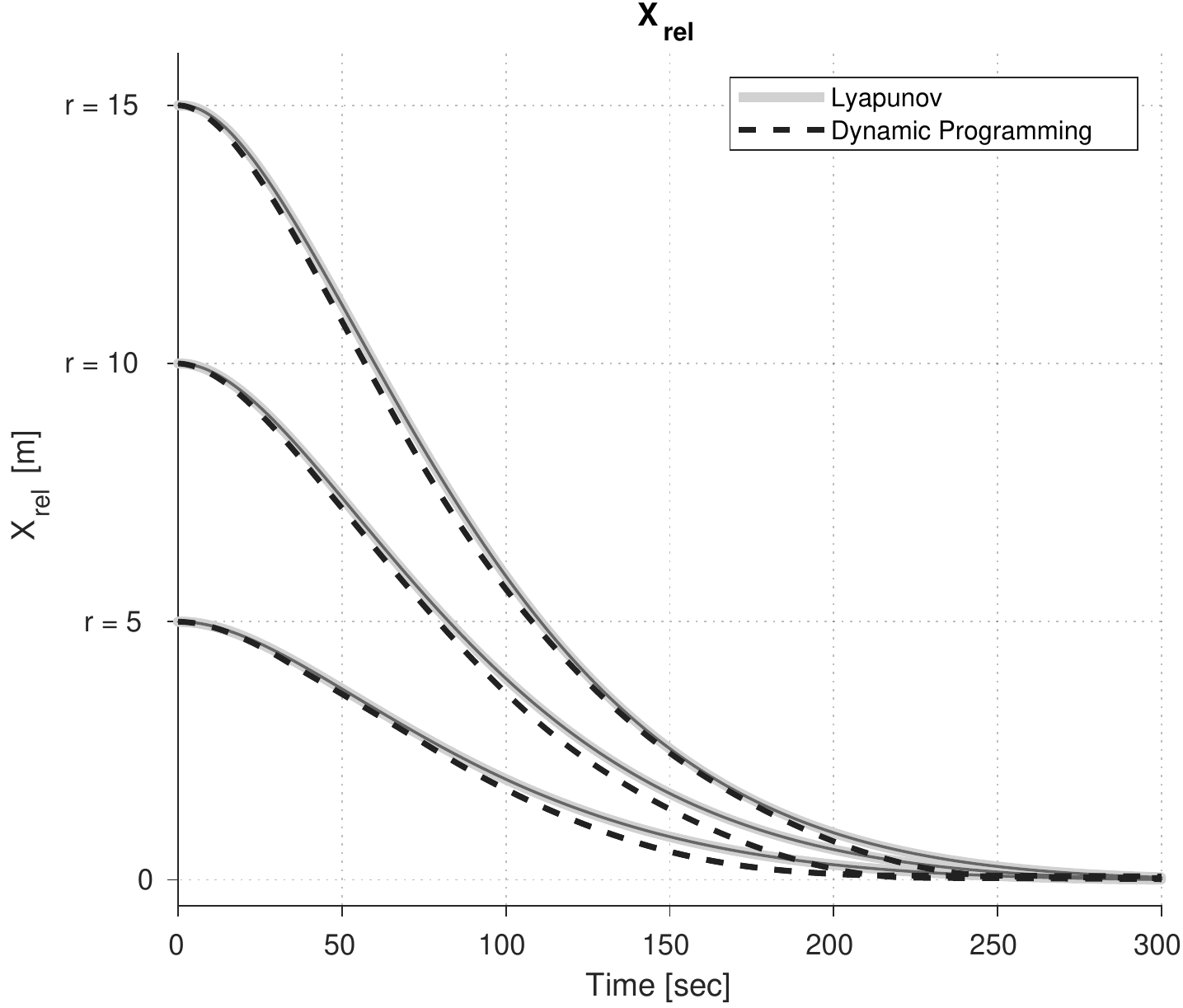}}
    \endminipage\hfill
    \minipage{0.32\textwidth}
    {\includegraphics[width=\linewidth]{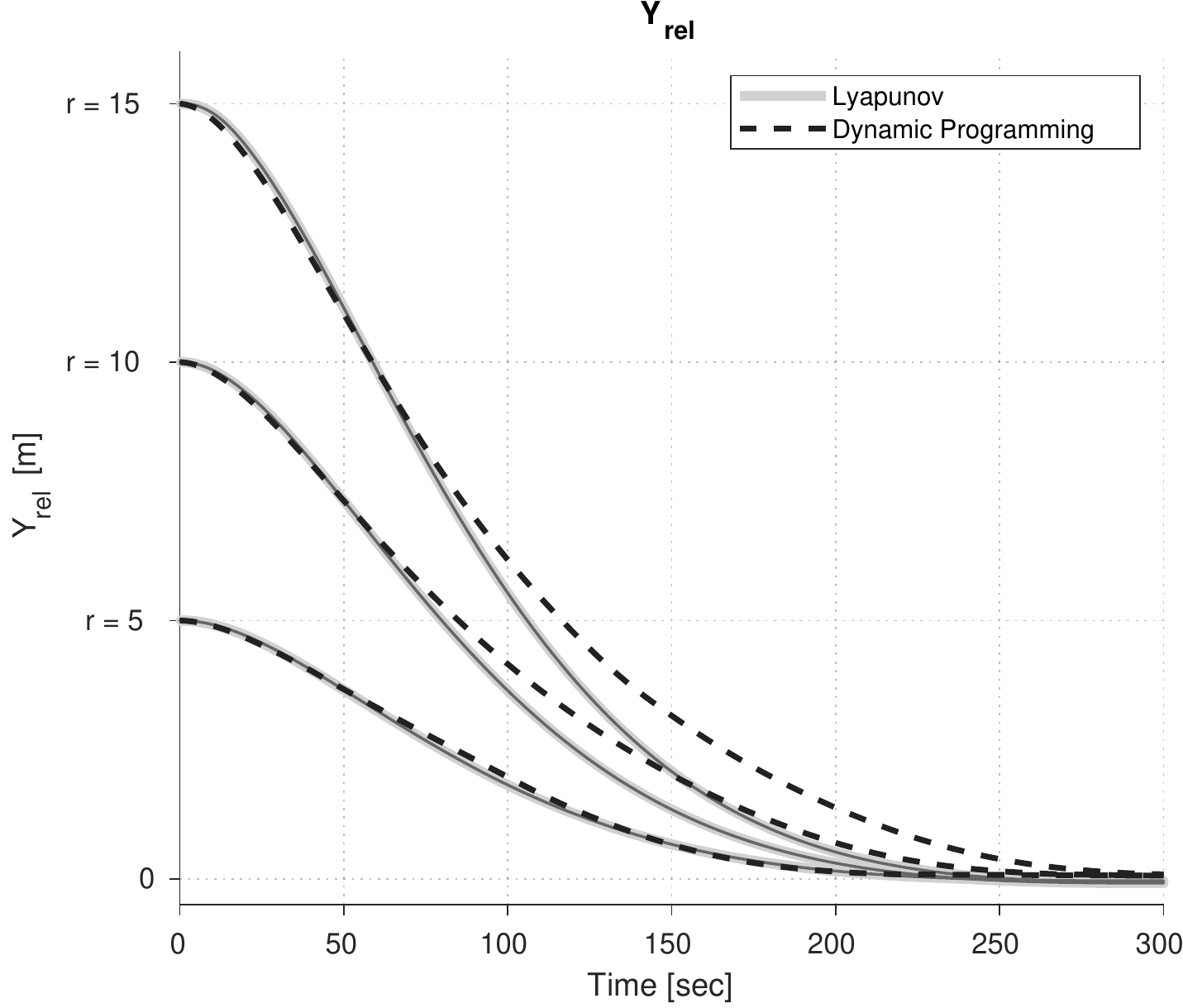}}
    \endminipage\hfill
    \minipage{0.32\textwidth}
    {\includegraphics[width=\linewidth]{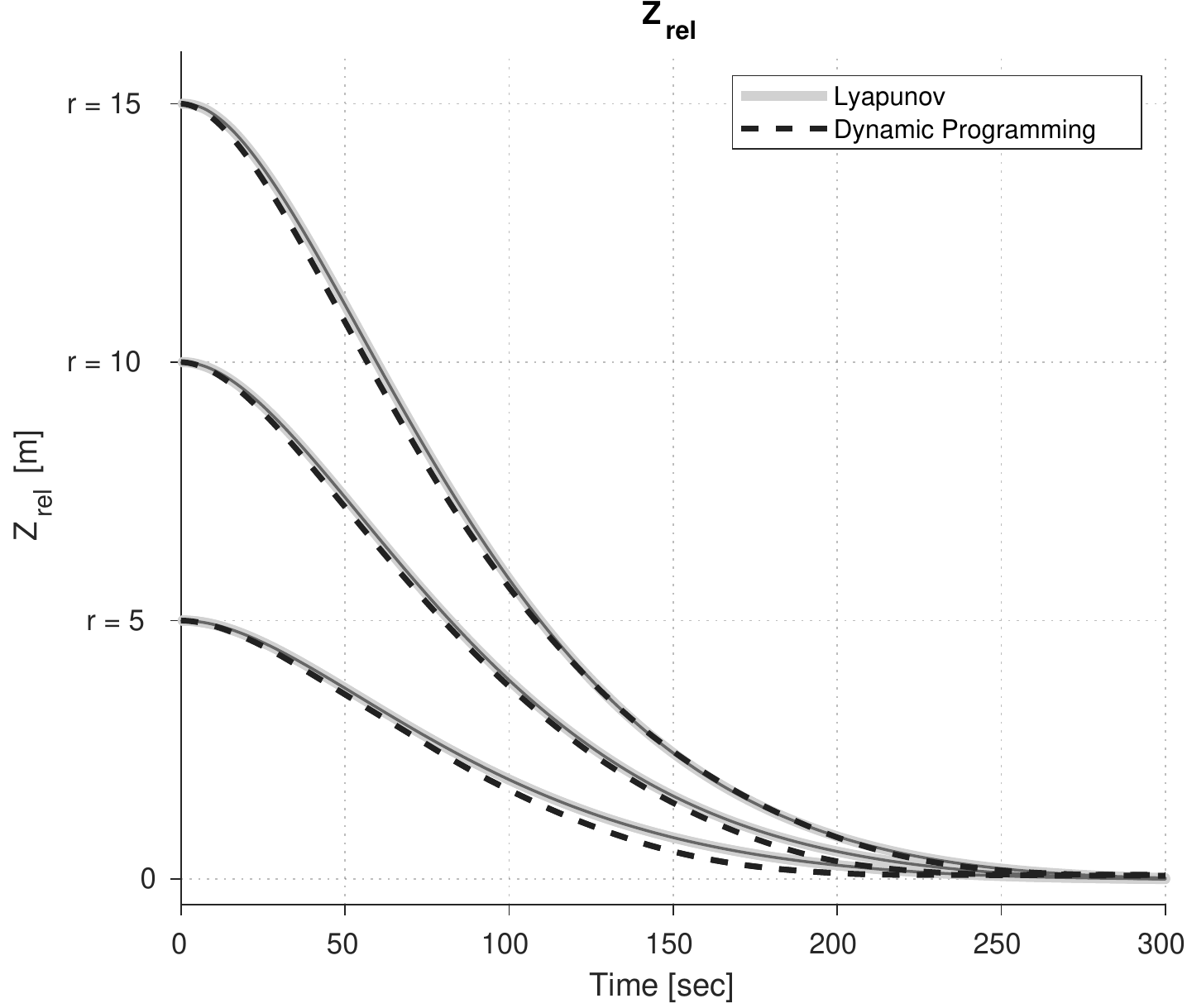}}
    \endminipage
    \caption{Comparison of the trajectories starting from various initial positions $\bm{\rho}_{0}=\left[r \ \ r \ \ r\right]^T$}
    \label{fig_3trajectories}
  
    \end{figure}

In view of the fact that the parameters in Table~\ref{tab:DPparams} and Table~\ref{tab:Lyapparams} were optimized to match the responses for given initial conditions, it is evident that if the simulations were carried out at different starting positions, both the settling times and impulses would be different. Thus, the variations in settling times and ratio of total impulses as a function of the initial position $\rho_0 = [r,r,r]^T$ for $5 \leq r \leq 15$ are analyzed and presented in \figurename~\ref{fig:maxsettle} and \figurename~\ref{fig:impulses}. Figure~\ref{fig_3trajectories} illustrates the trajectories of both control methods for the upper and lower bounds of the analyzed intial conditions.

It should be noted that total impulses are not calculated for the whole duration of simulations since excessive jittering in the Lyapunov-based control for $t \geq t_s$ aggravates its reported fuel consumption. Instead, first the highest settling time among all three channels is found (\figurename~\ref{fig:maxsettle}) and the total impulses are reported for up to 10 seconds after the highest settling time. Figure \ref{fig:maxsettle} shows that the settling time for dynamic programming varies with respect to initial position, whereas the settling time for the Lyapunov-based control follows the behavior of the second order reference models and is merely a function of the parameters $\left\{K_1,\ldots,K_4\right\}$. Additionally, since these parameters were tuned for the initial conditions of the fully-actuated scenario ($r=10$), they no longer comply with the constraints of Eq.~\eqref{eq:lyap_constr} for $r<10$, so the actual responses begin to detach from the reference models and the actual settling times increase accordingly.

\subsection{Simulation in Thruster-Failed Mode} \label{subsec:sim_fault}

In this section, we consider two cases with single thruster failures. In both cases, the spacecraft starts from the initial position $\bm{\rho}=\left[-10 \ -10 \ -10\right]^T$ with the initial angles and velocities set to zero. Thruster failure in different channels and approaching directions are presented, and these two cases are representative of other possible scenarios of approach under failure as well. The parameters used for the dynamic programming controllers in thruster failure case studies are increased back to the values listed in Table~\ref{tab:dpfault} as there is no longer an obligation to match the responses to Lyapunov-based control. The controller of the faulty channel uses a more relaxed set of parameters to take into consideration the fact that achieving very small steady-state errors for both the position and attitude along the faulty channel might be difficult with thruster control. 

{\begin{table}[htbp]
    \centering
    \caption{Dynamic programming parameters in the fault case studies}
    \begin{tabular}{lcc}
        \toprule
        \multirow{2}{*}{Parameters} & \multicolumn{2}{c}{Value}\\
        \cmidrule{2-3}
        & Position Control & Attitude Control \\
        \midrule 
    $Q_{\textrm{normal\:channels}}$ & $\left[\begin{matrix}1&0\\0&200\\\end{matrix}\right]$ &$\left[\begin{matrix}1&0\\0&5\\\end{matrix}\right]$ \\
        $R$ & 1 & 1 \\ 
    $ Q_{\textrm{faulty\:channel}} $ & $\left[\begin{matrix}0.01&0\\0&5\\\end{matrix}\right]$ &$\left[\begin{matrix}1&0\\0&5\\\end{matrix}\right]$ \\
        \bottomrule
    \end{tabular}
    \label{tab:dpfault}
\end{table}}

\begin{figure}[!tp]
    \centering
    {\includegraphics[width=0.82\columnwidth]{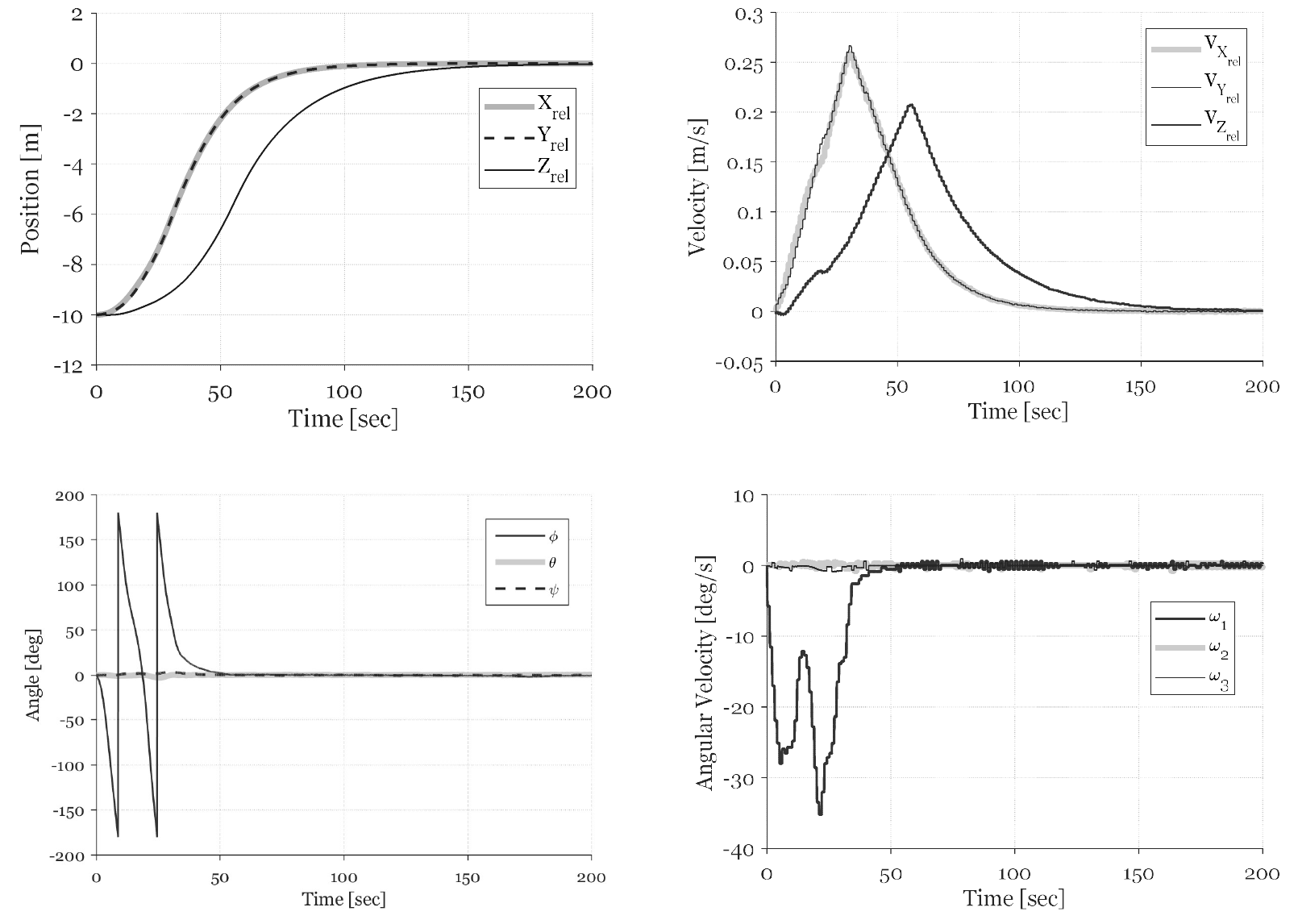}}
    \caption{Simulation results of the 6-DOF maneuver for thruster failure case study 1}
    \label{fig:simfault1}
  \end{figure}

\begin{figure}[!htp]
    \centering
    {\includegraphics[width=0.65\columnwidth]{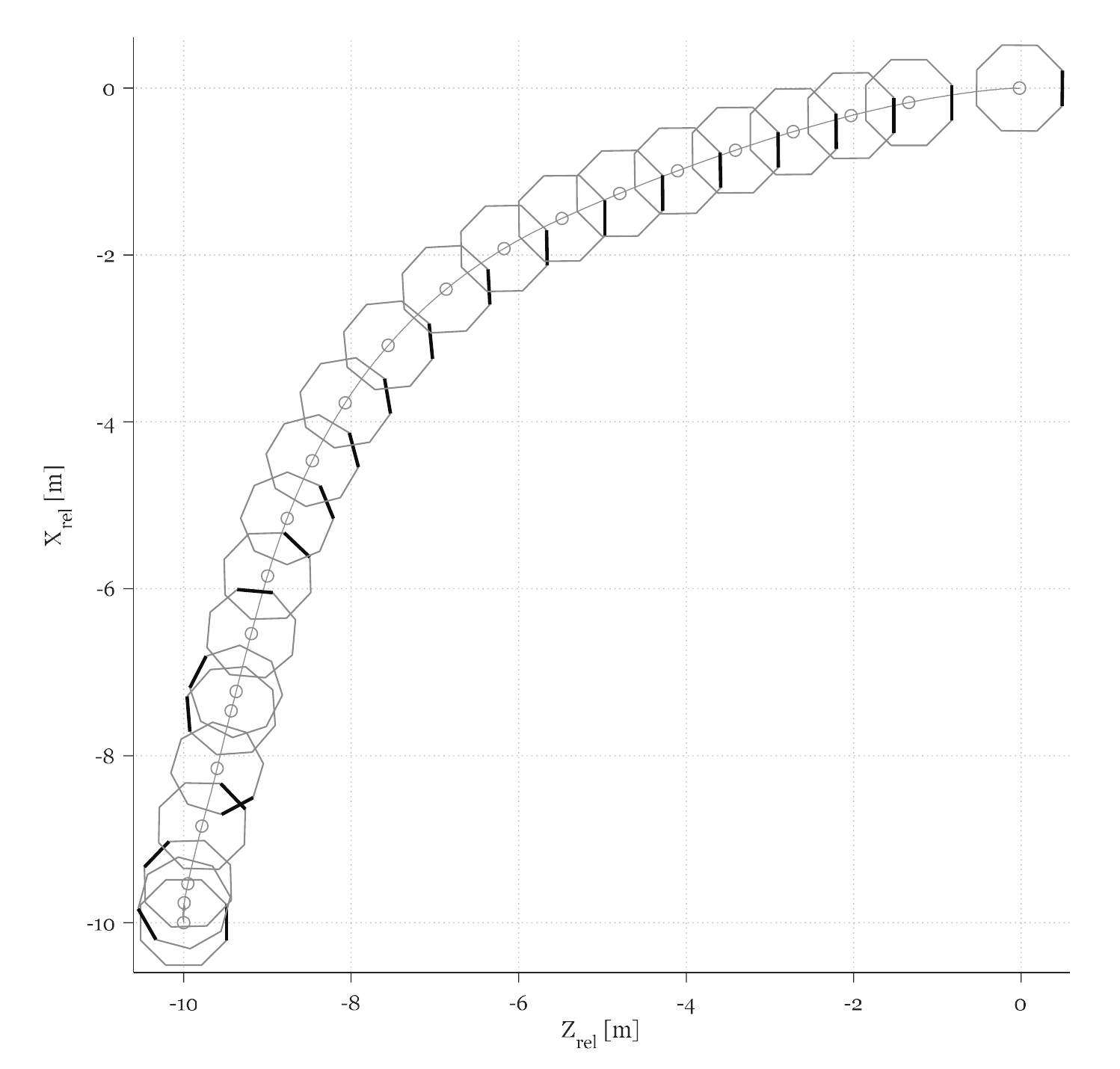}}
    \caption{View of the XZ-plane in thruster failure case study 1. The front side of satellite is drawn with thicker line to indicate its orientation}
    \label{fig:xz}
  \end{figure}

In the first case, a thruster that generates forces in the positive direction of the z-channel ($u_5$) fails. As a result of this failure, whenever the remaining thruster in the same direction fires (e.g., to move the satellite from the initial position to the target on the origin), the satellite gains both linear and rotational acceleration simultaneously.
 The simulation results for this case are shown in \figurename{~\ref{fig:simfault1}} to \figurename~\ref{fig:xz}. It is observed that the satellite gains rotational velocity at the beginning of the maneuver and flips twice for controlling the position, and ultimately manages to zero out its position and attitude smoothly because both thrusters that generate forces in the negative direction of the z-channel are operational. Since the satellite is approaching the target from the side with two operative thrusters, the weighting variable of the control allocator is constant ($W_{mf}=0.93$) throughout the maneuver.

  \begin{figure}[!htp]
    \centering
    {\includegraphics[width=0.85\columnwidth]{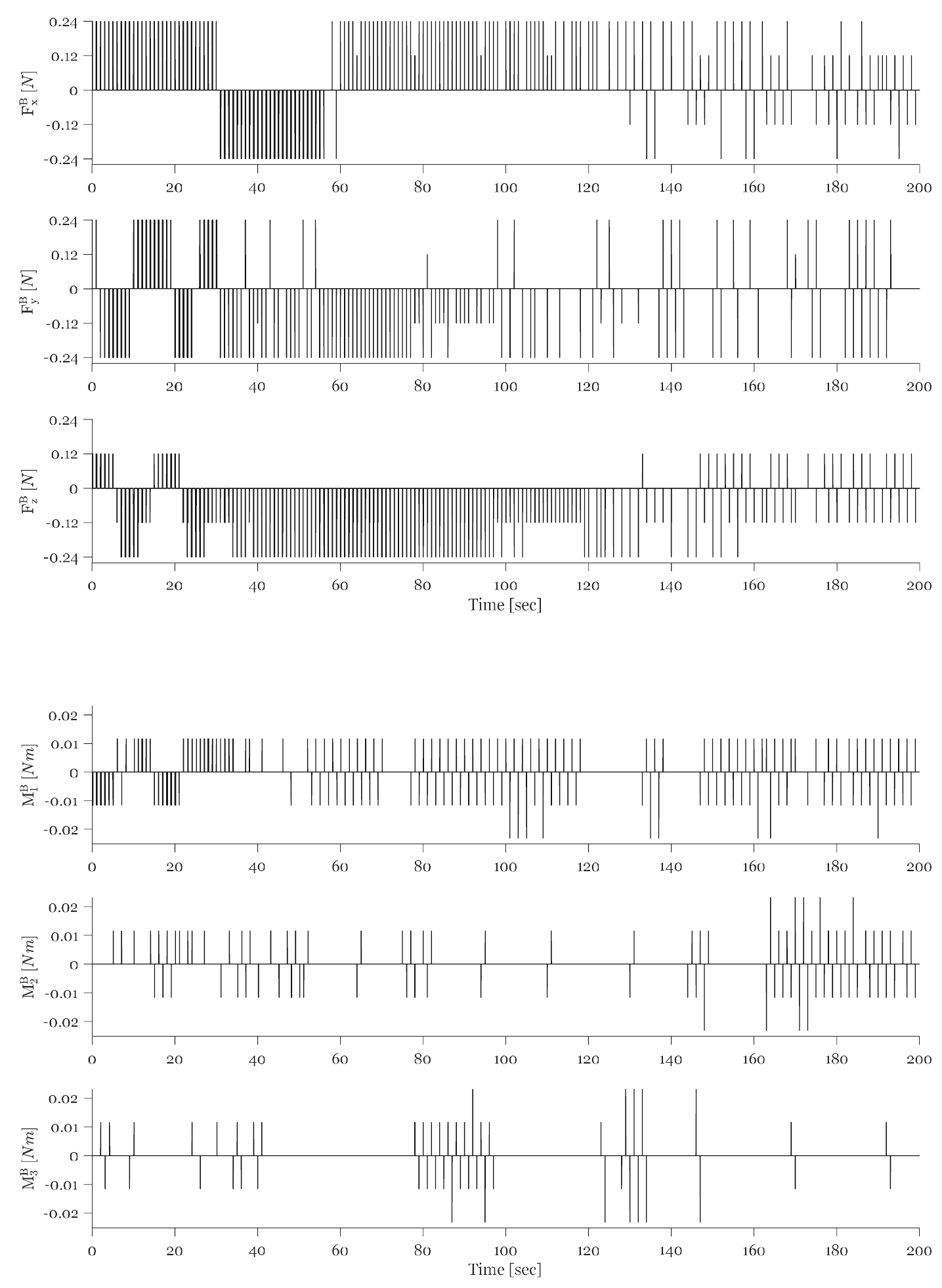}}
    \caption{Thruster-generated body forces and torques for thruster failure case study 1}
    \label{fig:FMfault1}
  \end{figure}
  
  \begin{figure}[!htp]
    \centering
    {\includegraphics[width=0.85\columnwidth]{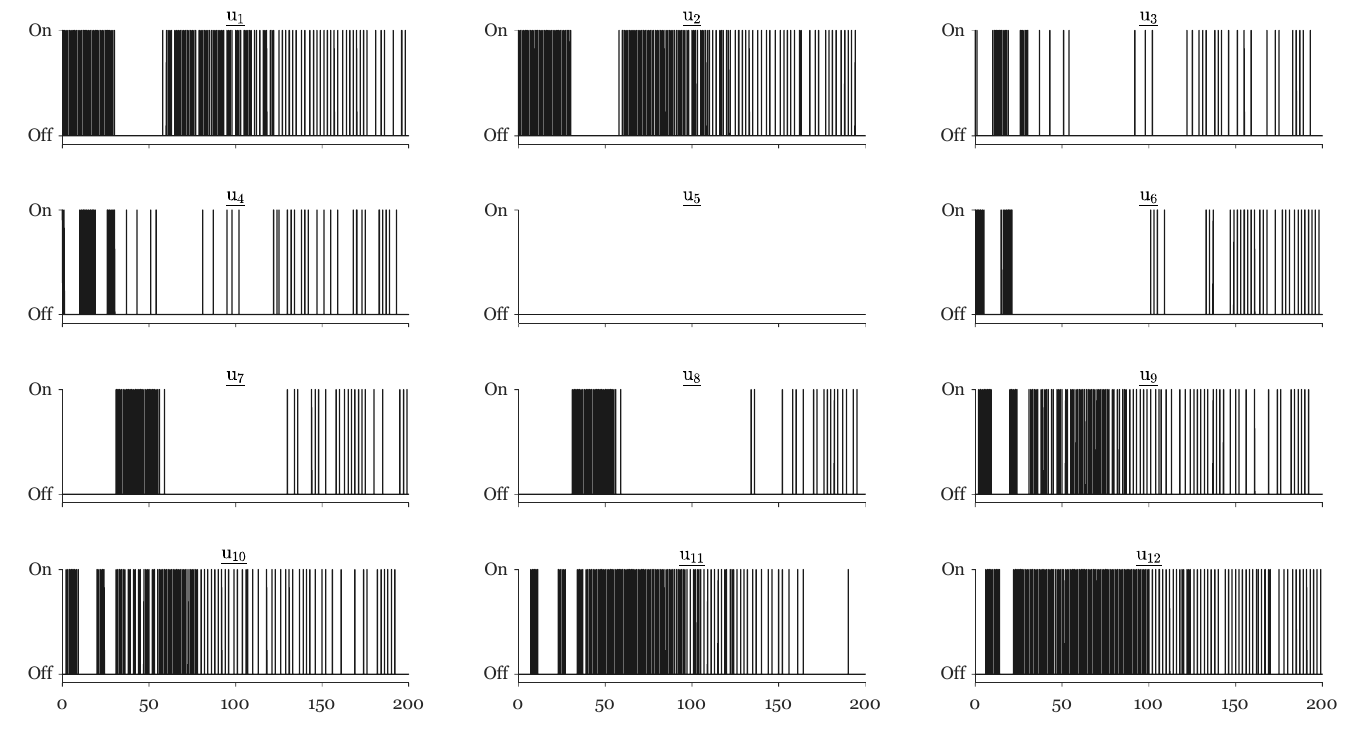}}
    \caption{Thruster firings for thruster failure case study 1}
    \label{fig:thrfault1}
  \end{figure}
  
\begin{figure}[!htp]
    \centering
    {\includegraphics[width=0.9\columnwidth]{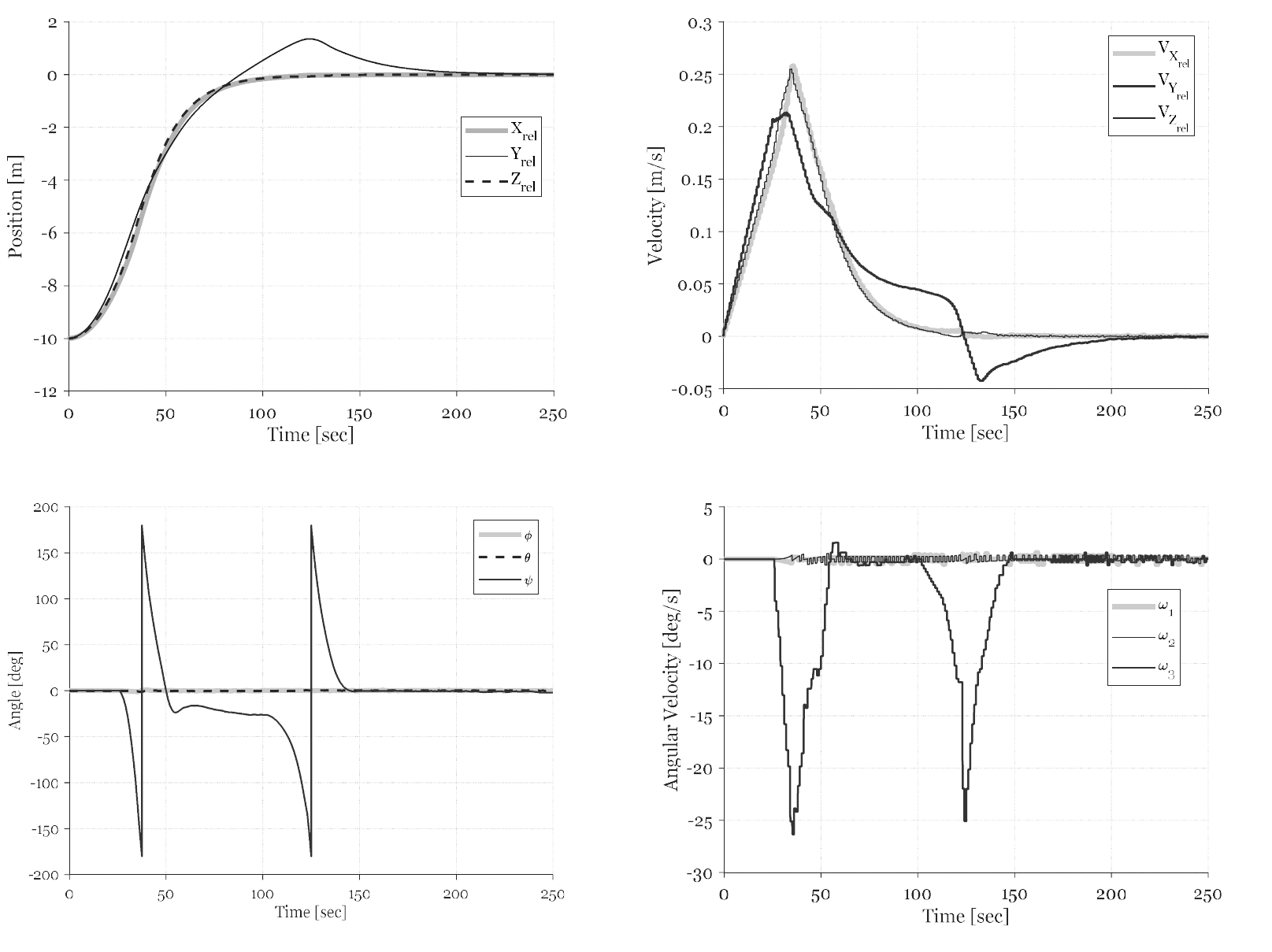}}
    \caption{Simulation results of the 6-DOF maneuver for thruster failure case study 2}
    \label{fig:simfault2}
  \end{figure}

\begin{figure}[!htp]
    \centering
    {\includegraphics[width=0.65\columnwidth]{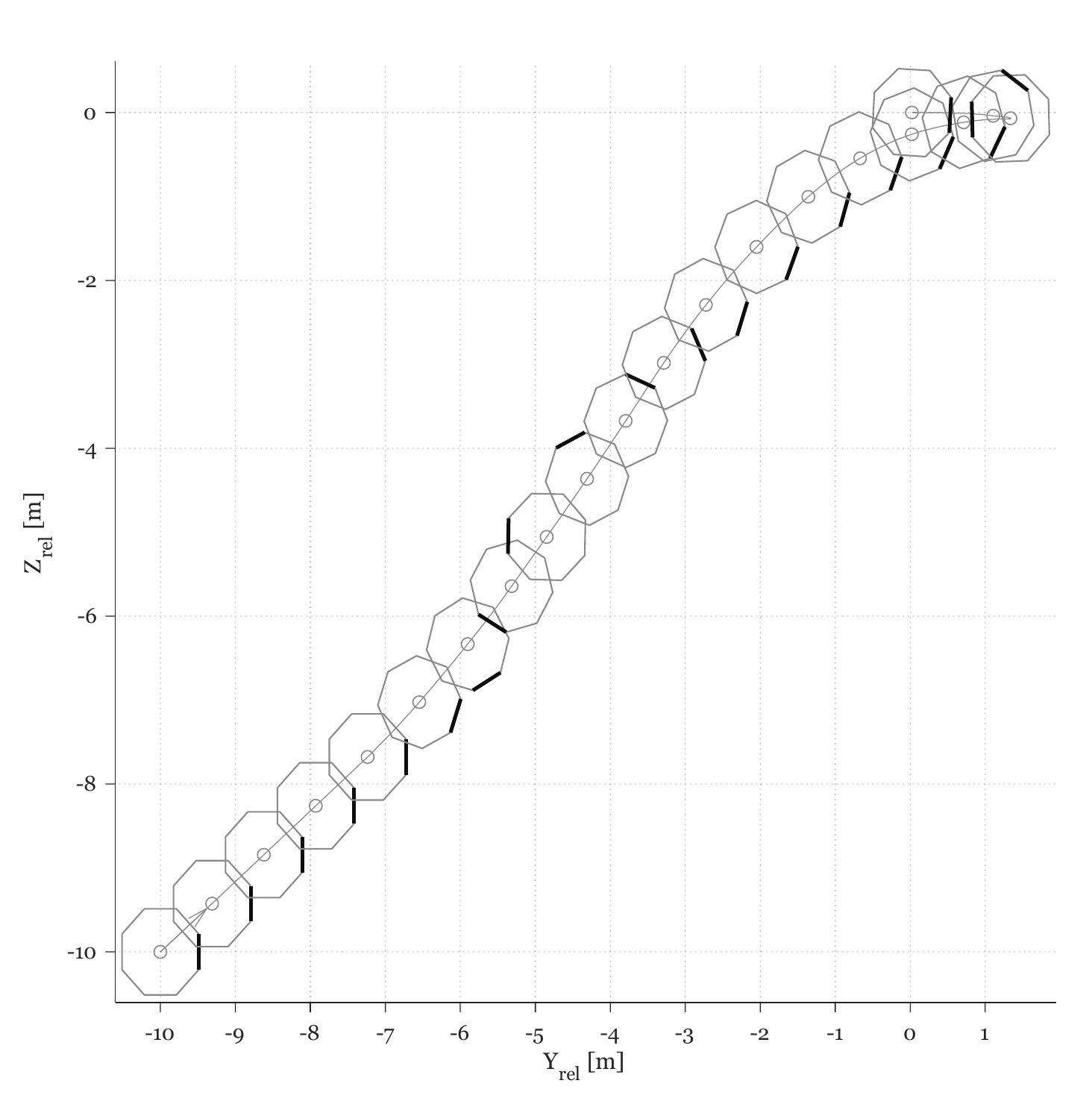}}
    \caption{View of the ZY-plane for thruster failure case study 2}
    \label{fig:zy}
  \end{figure}

In the next case, the satellite is controlled when a thruster in the negative direction of the y-channel ($u_{10}$) becomes inoperative. The simulation results for this case study are shown in \figurename~\ref{fig:simfault2} to \figurename~\ref{fig:zy}. This case has been given more attention in previous works \citep{Tavakoli2018,Pong2010} because it is fundamentally more difficult to control. This time, the remaining thruster in the same direction is required to provide negative acceleration to slow down and eventually stop the satellite on the target. However, activating a single thruster for this purpose is going to unwantedly spin the satellite as well. After the satellite first approaches the target~($t \approx 50 \, s$), since the control allocator does not find a thrusting strategy that provides negative acceleration and positive moment simultaneously, thrusting is only performed to keep the attitude stable while the satellite drifts to the other side of the target, after which it successfully controls both the position and attitude back to the target using both of its operative thrusters, as in the former case study.

  \begin{figure}[!htp]
    \centering
    {\includegraphics[width=0.83\columnwidth]{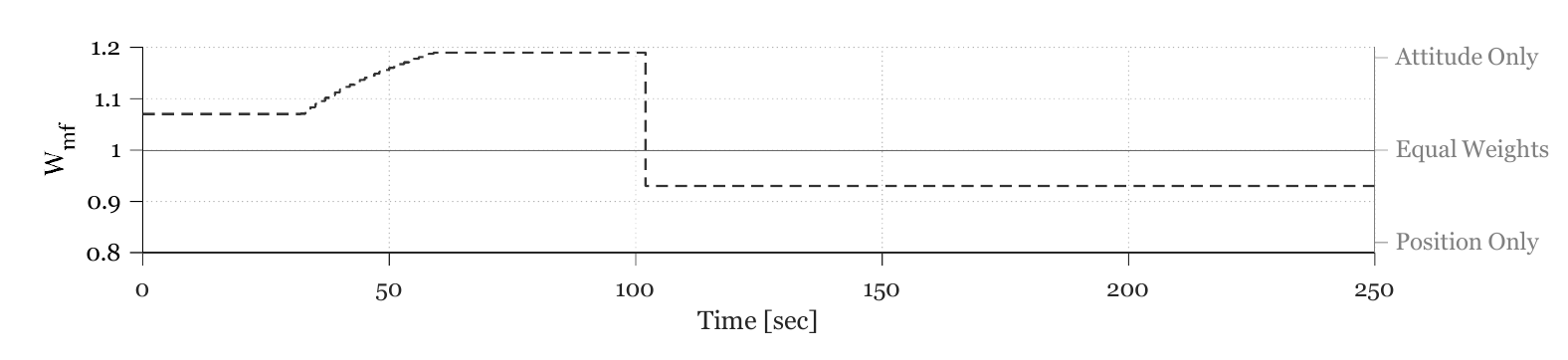}}
    \caption{Weighting variable of the control allocator for thruster failure case study 2}
    \label{fig:Wfault2}
  \end{figure}
  
  \begin{figure}[!htp]
    \centering
    {\includegraphics[width=0.85\columnwidth]{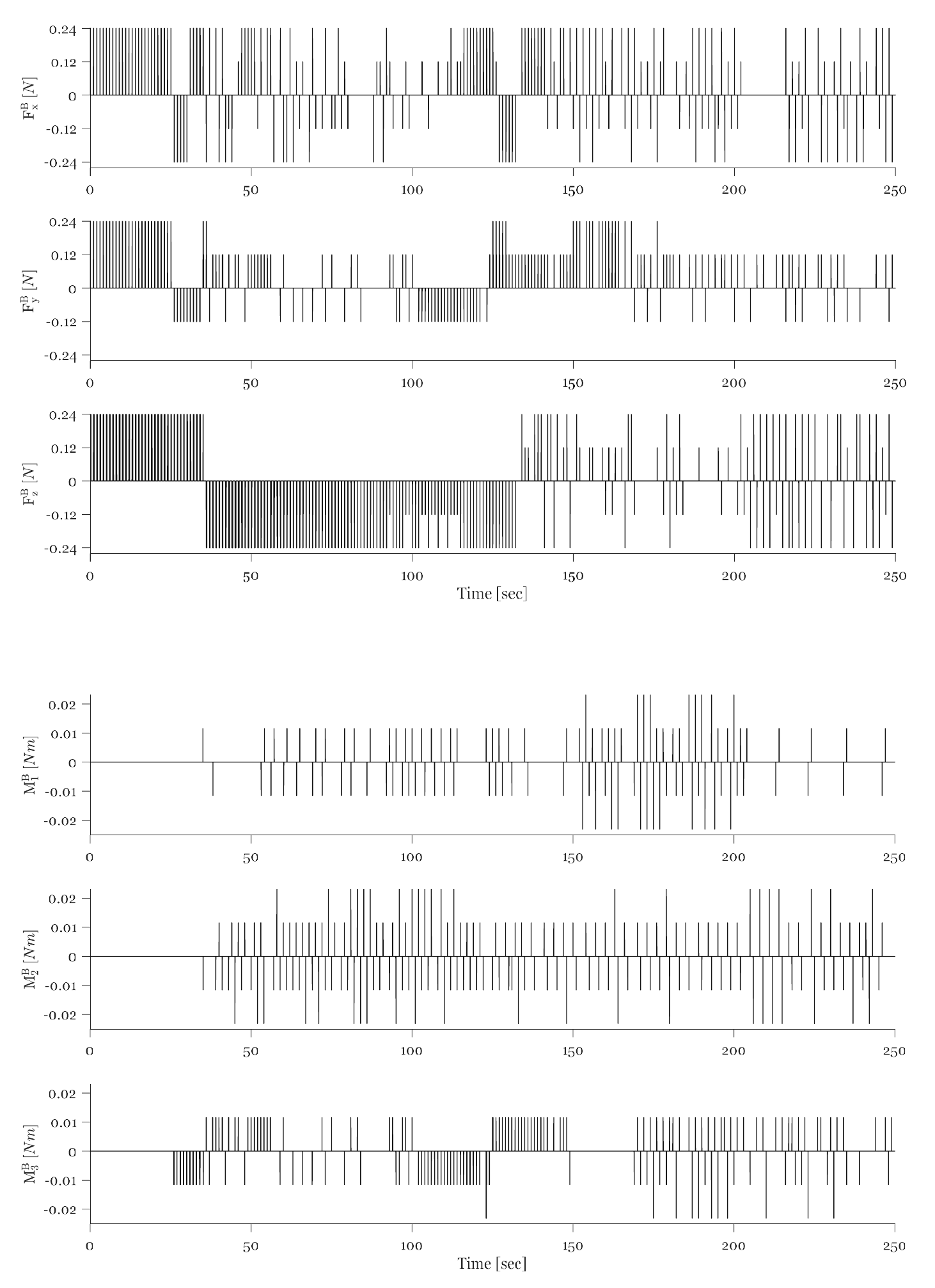}}
    \caption{Thruster-generated body forces and torques for thruster failure case study 2}
    \label{fig:FMfault2}
  \end{figure}
  
  \begin{figure}[!htp]
    \centering
    {\includegraphics[width=0.85\columnwidth]{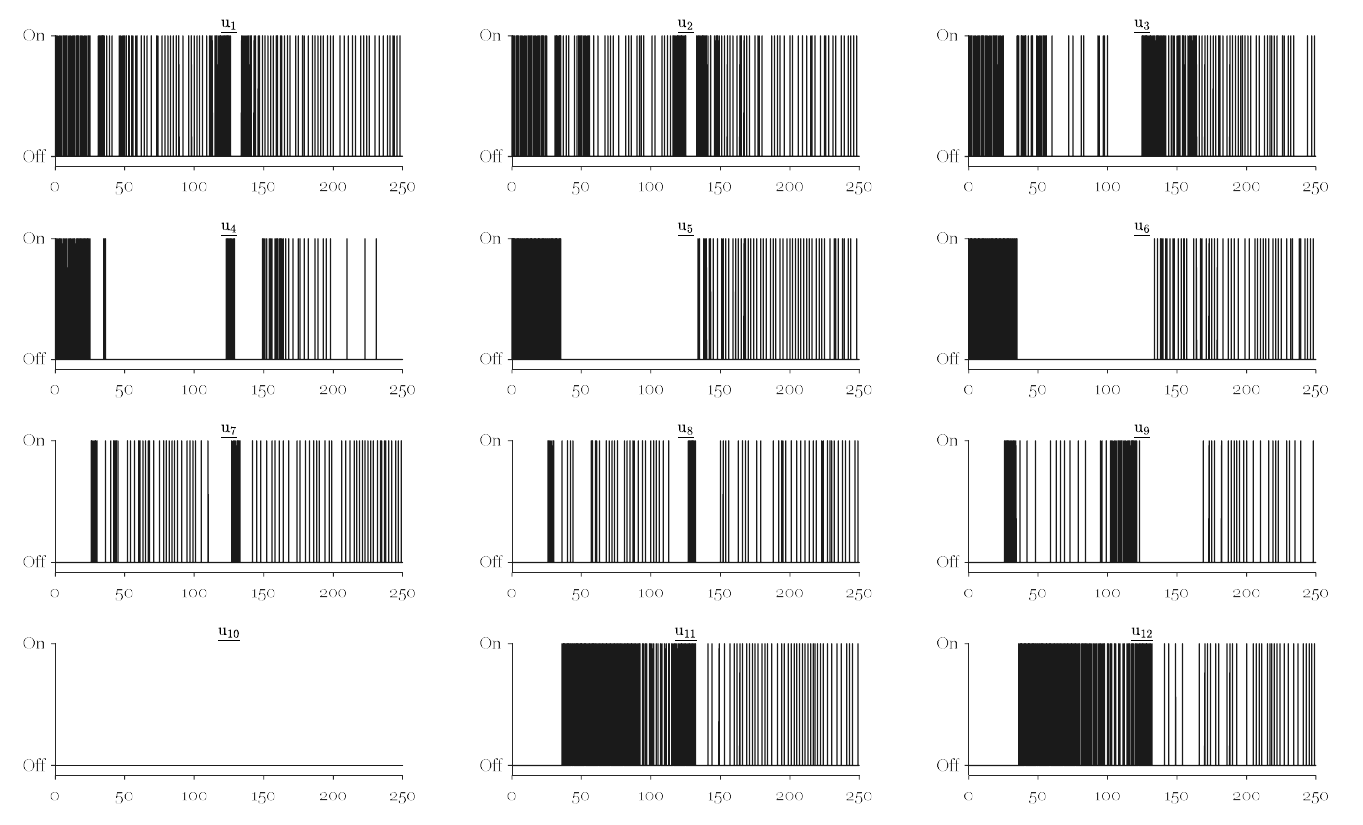}}
    \caption{Thruster firings for thruster failure case study 2}
    \label{fig:thrfault2}
  \end{figure}
  
\section{Conclusion} \label{sec:conclusion}
A control approach was developed in the present study for all-thruster spacecraft using the dynamic programming method. The analysis showed that dynamic-programming-based control was at least twice as more fuel efficient when compared to the Lyapunov-based control system that followed the same trajectory. Numerical sensitivity analysis for the fuel consumption as well as detailed results of both controls from different starting positions were also presented. For the under-actuated case, dynamic programming demonstrated a stable control that could not be achieved in previous studies.

The dynamic programming approach, by its nature, assumes perfect knowledge of the model and the states. Off-nominal conditions such as variations in thrust forces, unmodeled dynamics, output disturbances, and noises could significantly degrade the performance of the system. This is especially important for the under-actuated scenarios. Moreover, the developed control system requires information gathered by a fault detection and isolation mechanism. Future works could potentially expand the scope of the present analysis to a more robust control system.


\bibliography{references}

\end{document}